\documentclass[11pt]{article}
\usepackage{mydef2col}
\usepackage{kantlipsum,widetext}
\usepackage[T1]{fontenc}
\usepackage[autostyle=true]{csquotes}
\usepackage{todonotes}

\usepackage{autobreak}
\allowdisplaybreaks[4]

\graphicspath{{./Graphs/}}
\usepackage{tikz}
\usetikzlibrary{calc,decorations.markings}
\usetikzlibrary{arrows,patterns,positioning}

\usepackage[round]{natbib}
\def\Ito{It\^{o}'s }

\def\bth{{\bar{\theta}}}
\def\tth{{\Theta}}
\def\br{{\bar{r}}}
\def\hx{{\hat{x}}}

\newcommand{\Ind}{\mathbf{1}}
\newcommand{\LL}{\mathcal{L}}
\newcommand{\La}{\mathfrak{L}}

\newcommand{\yi}{y_{i}}
\newcommand{\yii}{y_{i+1}}

\def\bx{{\bar{x}}}
\def\bxi{{\bar{\xi}}}
\def\bOmega{{\bar{\Omega}}}
\def\bTheta{{\bar{\Theta}}}
\def\bchi{{\bar{\chi}}}

\def\bl{{\bar{l}}}
\def\by{{\bar{y}}}
\newcommand{\byi}{\bar{y}_{i}}
\newcommand{\byii}{\bar{y}_{i+1}}

\newcommand{\X}{\mathcal{X}}
\newcommand{\T}{\mathcal{T}}
\newcommand{\U}{\mathcal{U}}
\newcommand{\Y}{\mathcal{Y}}
\newcommand{\F}{\mathcal{F}}
\newcommand{\M}{\mathcal{M}}
\newcommand{\XI}{\X_{i}}
\newcommand{\TI}{\T_{i}}
\newcommand{\FI}{\F_{i}}
\newcommand{\UI}{\U_{i}}
\newcommand{\MI}{\M_{i}}

\newcommand{\Yi}{\Y_{i}}

\title{Multilayer heat equations: application to finance}
\def\thetitle1{Multilayer heat equations: application to finance}
\author{
\authorstyle{
Andrey Itkin{}
\textsuperscript{1}
Alexander Lipton{}
\textsuperscript{2}
and Dmitry Muravey
\textsuperscript{3}
}
\newline\newline
\textsuperscript{1}
\institution{Tandon School of Engineering, New York University, New York, USA} \\
\textsuperscript{2}
\institution{The Jerusalem School of Business Administration, The Hebrew University of Jerusalem, Jerusalem, Israel;} \\
\textsuperscript{\ \ }
\institution{Connection Science and Engineering, Massachusetts Institute of Technology, Cambridge, MA, USA} \\
\textsuperscript{3}
\institution{Moscow State University, Moscow, Russia}
}

\date{\today}
\begin{document}

\maketitle

\lettrineabstract{In this paper, we develop a Multilayer (ML) method for solving one-factor parabolic equations. Our approach provides a powerful alternative to the well-known finite difference and Monte Carlo methods. We discuss various advantages of this approach, which judiciously combines semi-analytical and numerical techniques and provides a fast and accurate way of finding solutions to the corresponding equations. To introduce the core of the method, we consider multilayer heat equations, known in physics for a relatively long time but never used when solving financial problems. Thus, we expand the analytic machinery of quantitative finance by augmenting it with the ML method. We demonstrate how one can solve various problems of mathematical finance by using our approach. Specifically, we develop efficient algorithms for pricing barrier options for time-dependent one-factor short-rate models, such as Black-Karasinski and Verhulst. Besides, we show how to solve the well-known Dupire equation quickly and accurately.  Numerical examples confirm that our approach is considerably more efficient for solving the corresponding partial differential equations than the conventional finite difference method by being much faster and more accurate than the known alternatives.}

%%%%%%%%%%%%%%%%%%%%%%%%%%%%%%%%%%%%%%%%%%%%%%%%%%%%%%%%%%%%
\vspace{0.5in}

\section*{Introduction} \label{layer}
The problem of solving partial differential equations (PDEs) with moving boundaries appears naturally in various areas of science and technology. As mentioned in \citep{kartashov2001}, such problems have been known in physics for a long time. They arise in several fields, such as (a) nuclear power engineering and safety of nuclear reactors; (b) combustion in solid-propellant rocket engines; (c) laser action on solids; (d) the theory of phase transitions (the  Stefan problem and the Verigin problem); (e) the processes of sublimation in freezing and melting; (f) in the kinetic theory of crystal growth; etc., see \citep{kartashov1999} and references therein. Analytical solutions to these problems often require rather sophisticated methods., which were actively developed by the Russian mathematical school in the 20th century starting from A.V.~Luikov, and then by B.Ya.~Lyubov, E.M.~Kartashov, and many others.

As applied to mathematical finance, one of these methods - the method of heat potentials (HP) - was actively utilized by A.~Lipton and his co-authors to solve various mathematical finance problems, see \citep{Lipton2001,LiptonPrado2020} and references therein. A complementary method of a generalized integral transform (GIT) is developed in \citep{CarrItkin2020jd,ItkinMuravey2020r,CarrItkinMuravey2020} to price barrier and American options in the semi-closed form. These authors studied the time-dependent Ornstein-Uhlenbeck (OU), Hull-White, CIR, and CEV models.  An extension of the method of heat potentials for the Bessel process called the method of Bessel potentials is developed by \citep{CarrItkinMuravey2020}, who also describe a general scheme of how to construct the potential method for any linear differential operator with time-independent coefficients. Finally, they also extended the method of generalized integral transform to the Bessel process. In all cases, a semi-analytical (or semi-closed form) solution means that first, one needs to solve a linear Volterra equation of the second kind. Then the option price is represented as a one-dimensional integral.

\citep{CarrItkin2020jd,ItkinMuravey2020r,CarrItkinMuravey2020} show that the new method is computationally more efficient than the existing ones, such as the backward and forward finite difference methods while providing better accuracy and stability. Also, the heat potential and GIT methods do not duplicate but rather complement each other. The former provides very accurate results for short maturities, and the latter for long maturities.

Even though many new problems have been solved in the above-cited papers, some of the financial models are hard to solve by using these methods directly. For instance, this is the case for the Black-Karasinski model, popular among practitioners. Another problem is the calibration of the local (or implied) volatility surface in various one-factor models. Almost all popular analytic and semi-analytical methods approach the solution of this problem by doing it term-by-term, which, doubtless, produces computational errors. For more details, see \citep{ItkinLocalVol} and references therein.

In this paper, we attack this class of problems (some of them unsolved in the semi-analytical form) by using another method, which we call the method of multilayer (ML) heat equation. An alternative approach is given in \citep{Dias2014}, where an innovative technique of recursive images is presented to obtain solutions to the transient diffusion equation in a $N$-layered material based on the superposition of Green functions for a semi-infinite material. The solution is initially built for a single layer over a substrate by constructing a sequential sum of reflected image functions. These functions are chosen to satisfy in sequence the boundary conditions, first at the front interface, then at the back interface, then again at the front interface, and so on until the added functions' magnitude becomes negligible.

Based on this so-called "1-layer" algorithm, the author also constructs a "2-layer" algorithm by sequential application of the "1-layer" algorithm first to layer 1, then to layer 2, then again to layer 1, and so on. The sequential application of the $N-1$ algorithm naturally leads to the $N$-layer algorithm. This scheme works for the first and second kind boundary conditions but does not apply to the case where there is a contact resistance between layers or the convective heat transfer at the end interfaces.

Note that this algorithm as applied to the local volatility calibration problem is similar to the approach used in \citep{LiptonSepp2011iv, ItkinLipton2017, ELVG, GLVG}.

Since the ML method splits the whole (possibly infinite) domain in the space variable into a sequence of sub-domains, one could extend it naturally to solving parabolic equations with coefficients being functions of time $t$ and location $x$. At every sub-interval, the corresponding parabolic operator could be either approximated by the operator with the space-homogeneous coefficients or, possibly, reduced to the heat equation by a series of transformations. After either approximation or reduction, the ML method can be applied.

Moreover, the method could be extended further to deal with non-linear volatility, drift, and killing term. Again, piecewise approximations of these terms lead to the parabolic equations at every sub-interval that could be transformed to the heat equation. Then, the application of the ML method solves the problem.

The main idea of this paper is to combine the ML method with the method of heat potentials \footnote{More general potential methods, e.g., the method of Bessel potentials, can also be used in such a scheme.} and the GIT method. Since both provide a semi-analytical solution for sub-interval problems, a combination of these solutions within the ML heat equation method results in the problem's full solution, expressed explicitly via one-dimensional integrals. At each layer, these integrals depend either on the yet unknown potential density (in the HP method) or on the solution gradient at the layer's boundaries (the GIT method). These unknown functions solve the interconnected systems of the integral Volterra equations of the second kind derived in the paper. Once this solution is found (either numerically or, sometimes, analytically), the whole problem is solved. Note that one can transform the system of integral equations to linear equations on a time-space grid, which is lower banded (in our case, block lower triangular). Therefore, the corresponding system can be solved with complexity $O(M^2 N)$ where $N$ is the number of layers, and $M$ is the number of time steps, see \citep{ItkinMuravey2020r} in more detail.

We also propose a particular construction of the layers' internal boundaries, which allows the representation of every integral in the Volterra equation as convolution. Applying the Laplace transform, we obtain a system of linear equations with a block-tridiagonal matrix (it contains four blocks). This system can be efficiently solved numerically (with complexity $O(N)$). In some cases, it can be solved analytically. After this system's solution is found, we use the Gaver-Stehfest method to compute the inverse Laplace transform, also with linear complexity in the number of layers $N$. This algorithm solves the system of the Volterra equations and thus the whole problem.

We illustrate these novel ideas by representing several significant financial problems in the form suitable for solving them by the ML method. These problems include pricing barrier options in the time-dependent Black-Karasinski and modified Black-Karasinski (Verhulst) models, see \citep{ItkinLiptonMuravey}, as well as the solution of the Dupire equation. We also provide several numerical examples to demonstrate our method's high speed and accuracy compared with standard finite-difference  (FD) methods.

To the best of our knowledge, all the paper results are new and contribute to the existing financial and physics literature. It is interesting to note that our method is capable of solving similar problems that appear in medicine and biology in addition to finance. For instance, our technique is well-suited for studying (a) the growth of diffusive brain tumors, which considers the brain tissue's heterogeneity, \citep{Asvestas2014}; (b) the transdermal drug release from an iontophoretic system, \citep{Pontrelli2016}; (c) and many other similar problems. It is imperative to emphasize that our method allows solving the ML problems with time-dependent boundaries and time- and space-dependent diffusion coefficients. In contrast, the method of \citep{CarrMarch2018} and all other known approaches operate only with constant boundaries (possibly with time-dependent boundary conditions) and spatially piecewise constant diffusion coefficients. Moreover, their setting corresponds to one of our numerical examples in Section~\ref{results}. Since in \citep{CarrMarch2018} the solution is obtained by using spectral (eigenvector) series, while we apply the Laplace transform method, our approach is about 1000 times faster.

The rest of the paper is organized as follows. In Section~\ref{solveMHE} we construct the solution of the ML heat equation by using the method of heat potentials. In Section~\ref{GIT}, we solve this equation by using the GIT method. In Section~\ref{genBK} we describe the pricing of barrier options in the time-dependent Black-Karasinski (BK) model and also in our modification of this model, which was introduced in \citep{ItkinLiptonMuravey} and is called the Verhulst model. In particular, we demonstrate how to reduce the pricing PDEs for both models to the ML heat equation. Also, in Section~\ref{genBK} we provide a generalization of this approach for some other models. In Section~\ref{Dupire} we apply the results of Section~\ref{sec:sigmax} to investigate the case of space-dependent volatility $\sigma(x)$ in conjunction with solving the Dupire equation. Section~\ref{solVolt} is dedicated to the solution of the Volterra equations. In particular, we describe a construction of the internal boundaries, which allows a transformation of the Volterra equations of the second kind to Abel equations. We solve the latter equations via the Laplace transform. Section~\ref{results} describes some numerical experiments with the ML method. The final section concludes.

\section{Solving the ML heat equation via the HP method} \label{solveMHE}

Let us consider the following initial-boundary problem
\begin{align} \label{L_Cauchy_problem1}
\LL u(\tau,x) &= 0, \qquad  (x,\tau) \in \Omega: \left[y^-(\tau), y^+(\tau) \right] \times \mathbb{R}_{+}, \\
u(0,x) &= f(x), \quad y^-(0) < x < y^+(0), \nonumber \\
u(\tau,y^-(\tau)) &= \chi^-(\tau), \qquad u(\tau, y^+(\tau)) = \chi^+(\tau). \nonumber
\end{align}
Here the operator $\LL$ is a partial differential operator of the parabolic type
\begin{equation}
\LL = -\fp{}{\tau} + \fp{}{x} \left( \sigma^2(\tau,x) \fp{}{x} \right) + \mu(\tau,x) \fp{}{x} + \nu(\tau,x),
\end{equation}
\noindent $\sigma(\tau,x), \mu(\tau,x), \nu(\tau,x)$ are some known functions,  $\Omega$ is the spatial-temporal domain with curvilinear temporal boundaries, and $\chi^-(\tau), \chi^+(\tau)$ are known functions of time (the boundary conditions).

Similar to \citep{ItkinMuravey2020r}, we represent the solution in the form
\begin{equation} \label{q1}
u(x,\tau) = q(x,\tau) + \int_{y^-(0)}^{y^+(0)} f(\xi){\cal G}(x, \xi, \tau) d\xi,
\end{equation}
\noindent where ${\cal G}(x, \xi, \tau)$ is Green's function of the problem. Then the function $q(x,\tau)$ solves a problem similar to \eqref{L_Cauchy_problem1} but with the homogeneous initial condition
\begin{align} \label{L_Cauchy_problem}
\LL q(\tau,x) &= 0, \qquad  (x,\tau) \in \Omega: \left[y^-(\tau), y^+(\tau) \right] \times \mathbb{R}_{+}, \\
q(0,x) &= 0, \quad y^-(0) < x < y^+(0), \nonumber \\
q(\tau,y^-(\tau)) &= \chi^-(\tau) - \int_{y^-(0)}^{y^+(0)} f(\xi) {\cal G}(y^-(\tau), \xi, \tau) d\xi = \phi^-(\tau), \nonumber \\
q(\tau, y^+(\tau)) &= \chi^+(\tau) - \int_{y^-(0)}^{y^+(0)} f(\xi) {\cal G}(y^+(\tau), \xi, \tau) d\xi = \phi^+(\tau). \nonumber
\end{align}

If the Green function ${\cal G}(x, \xi, \tau)$ is known, the problem in \eqref{L_Cauchy_problem} can be solved via the HP method, \citep{ItkinMuraveyDB}. Otherwise, one can apply the ML method as this is described below.

To use the ML method, suppose the domain $\Omega$ could be split into $N$ layers: $\Omega = \bigcup_{i=1}^N \Omega_i$, where each layer is a curvilinear strip
\begin{align} \label{Omega_i_def}
\Omega_i &= [y_{i}(\tau), y_{i+1}(\tau)] \times \mathbb{R}_{+}, \quad y_{i}(\tau) < y_{i+1}(\tau), \quad \forall \tau >0, \quad \forall i = 1,\ldots,N, \\
y_1(\tau) &= y^-(\tau), \quad y_{N+1}(\tau) = y^+(\tau). \nonumber
\end{align}

Let us seek for the solution of the problem \eqref{L_Cauchy_problem} in the form
\begin{equation} \label{u_sum_repr}
u(\tau,x) = \sum_{i = 1}^{N} u_{i}(\tau,x) {\mathbf 1}_{x-\yi(\tau)} {\mathbf 1}_{\yii(\tau)-x}, \qquad
{\mathbf 1}_x =
\begin{cases}
1, & x \ge 0 \\
0, & x < 0,
\end{cases}
\end{equation}
\noindent and request that both $u(\tau,x)$ and its flux are continuous functions of $x$ \footnote{These conditions are natural in physics if by $u(t,x)$ we assume, e.g., the temperature and interpret $\sigma^2 \partial_x u(t,x)$ as the heat flux. Therefore, it is standard to require continuity of the heat flux rather than the first derivative $\partial_x u(t,x)$, \citep{ahtt5e}}. Using these conditions at every boundary $y_i(\tau), \ i=2,\ldots,N$ together with the boundary conditions yields the following system of equations
\begin{align} \label{matching_layers}
u_{i}(\tau,\yii(\tau)) &= u_{i+1}(\tau,\yii(\tau)),  \\
\sigma_i^2(\tau,\yii(\tau)) \fp{u_i}{x}\Bigg|_{x = \yii(\tau)} &=
\sigma_{i+1}^2(\tau,\yii(\tau)) \fp{ u_{i+1}}{x}\Bigg|_{x = \yii(\tau)}, \quad i = 1,\ldots,N-1, \nonumber \\
u_{1}(\tau,y^-(\tau)) &= \chi^-(\tau), \qquad u_{N +1}(\tau,y^+(\tau)) = \chi^+(\tau). \nonumber
\end{align}
The first condition means a continuity of the function $u$ at every boundary $y_i, \ i=1,\ldots,N-1$. The second condition is a continuity of the heat flux at the same boundary. The last line follows from the boundary conditions in \eqref{L_Cauchy_problem}.

Also, let us define the operator $\LL$ for the whole domain $\Omega$ as follows
\begin{equation} \label{L_def}
\LL  = \sum_{i = 1}^{N} \LL_{i} {\mathbf 1}_{x-\yi(\tau)} {\mathbf 1}_{\yii(\tau)-x},
\end{equation}
\noindent where
\begin{equation} \label{LLi_def}
\LL_i =  - \fp{}{\tau} + \fp{}{x} \left(\sigma^2_i(\tau, x)\fp{}{x} \right) + \mu_i(\tau,x) \fp{}{x} + \nu_i(\tau,x),
\end{equation}
\noindent and $\LL_i u_i = 0$.

The idea of the ML method is to assume that Green's function ${\cal G}_i(x, \tau|\xi, s)$ associated with the operator $\LL_i$ can be obtained in the closed form. For an arbitrary dependencies $\sigma_i(\tau,x), \mu_i(\tau,x), \nu_i(\tau,x)$ this is not the case, but for various specific forms of these functions this can be done. For instance, when $ \mu(\tau,x) = \nu(\tau, x) = 0$ and $\sigma(\tau,x) = \sigma(\tau)$ or $\sigma(\tau,x) = \sigma(x)$, etc., \citep{Polyanin2002}.  Otherwise, the functions $\sigma_i(\tau,x), \mu_i(\tau,x), \nu_i(\tau,x)$ can be approximated at every layer, e.g., by piecewise constant or linear function in $x$ and an arbitrary function of $\tau$, or by piecewise constant functions in $\tau$ and piecewise linear functions in $x$, etc. These approximations make the ML method somewhat similar to the FD method, however, with some critical distinctions, see Section~\ref{Discus}.

It is important to mention, that the operator $\LL_i$ in \eqref{LLi_def}, while natural for physics where a divergent form of the parabolic equation (e.g., the heat equation) is commonly accepted, is just rarely used in mathematical finance. Instead, in finance it is natural to consider a non-divergent (non-conservative) form, which for the heat equation reads
\begin{equation} \label{LLi_defND}
\LL_i =  - \fp{}{\tau} + \sigma^2_i(\tau,x)\sop{}{x}.
\end{equation}
Obviously, when $\sigma_i = \sigma_i(\tau), \forall i$, i.e. $\sigma_i(\tau, x)$ is a straight line at given $\tau$, both operators in \eqref{LLi_def} and \eqref{LLi_defND} coincide. However, if one solves \eqref{LLi_defND} by the ML method, it can be unclear what continuity condition should be used. As shown in \citep{Lejay2006}, for the divergent heat/diffusion equation with drift this condition remains the same, i.e. this is a continuity of flux over the boundary. Obviously, a non-divergent heat equation can be represented in this form, i.e. the divergent diffusion part plus drift. Therefore, the continuity condition is still represented by the equality of fluxes over the boundary, but the equation now includes an extra drift term.

As applied to the ML method, this can be seen as follows. Suppose we apply the ML method to some parabolic equation, and approximate all coefficients in the drift and killing terms by piecewise constant function at every interval. Then, at the $i$-th interval this equation reads
\begin{equation}
\fp{u_i(\tau,x)}{\tau} = \sigma^2(\tau, x) \sop{u_i(\tau,x)}{x} + \alpha_i \fp{u_i(\tau,x)}{x} + \beta_i u_i(\tau,x),
\end{equation}
\noindent where $\alpha_i = const, \ \beta_i=const, \ i=1,N$. By transforming it to a divergent form we obtain
\begin{equation} \label{exEq}
\fp{u_i(\tau,x)}{t} = \fp{}{x} \left(\sigma^2(\tau,x) \fp{u_i(\tau,x)}{x}\right) + \left[ \alpha_i  - 2 \sigma(\tau,x) \sigma_x(\tau,x)\right] \fp{u_i(\tau,x)}{x} + \beta_i u_i(\tau,x).
\end{equation}
Hence, again, the continuity condition for this equation is given in \eqref{matching_layers}. Further, \eqref{exEq} by a series of transformations can be reduced to a non-divergent heat equation in \eqref{LLi_defND}. Accordingly, these transformations should be applied to \eqref{matching_layers} as well to obtain the correct continuity conditions.

When the external boundaries are constant, i.e. $y^-(t) = \chi^-(t) = const, \ y^+(t) = \chi^+(t) = const$ one may use an alternative where a non-divergent  heat equation can be reduced to a divergent one by a change of variables $x \mapsto y = g(x)$, where $g(x)$ is some function which depends on $\sigma^2(x)$. In more detail this is shown in Appendix~\ref{divA}. Accordingly, the operator \eqref{LLi_defND} transforms to
\begin{equation} \label{LLi_defNDy}
\LL_i =  - \fp{}{\tau} + \fp{}{y} \left(\Xi^2_i(\tau,y)\fp{}{y} \right),
\end{equation}
\noindent where $\Xi^2(\tau,y)$ is a new diffusion coefficient, which can be expressed via $\sigma(\tau,x)$, again see Appendix~\ref{divA}.

In what follows, we provide our analysis for \eqref{LLi_def}; \eqref{LLi_defND} can be analyzed similarly, as explained above.
For simplicity and without loss of generality, we give an exposition of the HP method assuming $\mu_i(\tau,x) = \nu_i(\tau,x) = 0$, and either $\sigma_i(\tau,x) = \sigma_i(\tau)$, or $\sigma_i(\tau,x)$ is a piecewise constant function of $x$ for every layer.  In this case each equation $\LL_i u_i = 0$  by some change of variables $\tau \mapsto \bar{\tau}, x \mapsto \bar{x}$ can be transformed to the heat equation \eqref{heat} with $\sigma^2_i(\bar{\tau},{\bar x}) = 1$\footnote{Of course, there exist other possible representations of $\sigma_i$ which give rise to the heat equation, or e.g., to the Bessel equation, \cite{CarrItkinMuravey2020}.}, \citep{Polyanin2002}, and the corresponding Green function $G(\bar{x},\xi,\bar{\tau})$ reads
\begin{equation} \label{green}
G(\bar{x},\xi,\bar{\tau}) = \frac{1}{2\sqrt{\pi \bar{\tau}}} e^{-\frac{(\bar{x}-\xi)^2}{4\bar{\tau}}}.
\end{equation}

Also, these transformations modify the boundary $y(\tau) \mapsto \bar{y}(\bar{\tau})$.  Some examples of such transformations are presented in Section~\ref{secApp}. In Appendix~\ref{App2} we also provide some recipes on how to proceed if one needs to generalize this approach by considering a general case $\sigma = \sigma(\tau,x)$.

To the end of this section, for easiness of reading let us drop the bar over new variables. Now, following the general idea of the method of heat potentials for pricing double barrier options, \citep{ItkinMuravey2020r,CarrItkinMuravey2020}, we represent each function $q_i(\tau,x)$ as
\begin{align} \label{u_i_HP_repres}
q_i(\tau_i,x) &= \int_0^{\tau_i} \Bigg\{ \Psi_i(k) \fp{G(x, \xi, \tau_i - k)}{\xi}\Bigg|_{\xi = \yi(k)} + \Phi_i(k) \fp{G(x,\xi, \tau_i - k)}{\xi}\Bigg|_{\xi = \yii(k)} \Bigg\}dk.
\end{align}
In \eqref{u_i_HP_repres} the second integral is a sum of two single layer potentials with the potential densities $\Psi_i(\tau)$ and $\Phi_i(\tau)$. By writing \eqref{u_i_HP_repres} we take into account that  according to \eqref{coef} and \eqref{trW}, the transformed time $\tau$ might differ for each interval, therefore, the notation $\tau_i$ is used. However, e.g.,  for the problem described in Section~\ref{genBK} all new times are equal, i.e. $\tau_i = \tau, \ i=1,\ldots,N+1$.

Since the domain $\Omega$ consists of $N$ layers, there are $2N$ unknown density functions $\Psi_i(\tau_i), \Phi_i(\tau_i), \ i=1,\ldots,N$. To determine them one need to plug the representation of $q_i(\tau_i,x)$ in \eqref{u_i_HP_repres} into \eqref{matching_layers}, and then solve thus obtained system of the integral Volterra equations of the second kind.

However, it is known, \citep{TS1963} that the integral in \eqref{u_i_HP_repres} for $x = \yi(\tau_i)$ and $x = \yii(\tau_i)$ is discontinuous, but with the finite value at $x = y_i(\tau_i) + \varepsilon, \ \forall i=1,\ldots,N+1$ when $\varepsilon \to 0$. Then, as shown in \citep{ItkinMuraveyDB}, \eqref{u_i_HP_repres} should be represented in the form
\begin{align} \label{solU}
q_i (\tau, y_{i}&(\tau)) = \frac{1}{2\sigma^2_i(y_{i}(\tau))} \Psi_i(\tau)  \\
&+\int_0^\tau \Bigg\{ \Psi_i(k) \fp{G(y_{i}(\tau), \xi, \tau - k)}{\xi}\Bigg|_{\xi = \yi(k)} + \Phi_i(k) \fp{G(y_{i}(\tau),\xi, \tau - k)}{\xi}\Bigg|_{\xi = \yii(k)} \Bigg\}dk, \qquad \quad \tau = \tau_i,\nonumber \\
q_i (\tau,y_{i+1}&(\tau)) = -\frac{1}{2\sigma^2_i(y_{i+1}(\tau))} \Phi_i(\tau) \nonumber \\
&+\int_0^\tau \Bigg\{ \Psi_i(k) \fp{G(y_{i+1}(\tau), \xi, \tau - k)}{\xi}\Bigg|_{\xi = \yi(k)} + \Phi_i(k) \fp{G(y_{i+1}(\tau),\xi, \tau - k)}{\xi}\Bigg|_{\xi = \yii(k)} \Bigg\}dk. \quad \tau = \tau_{i+1}.\nonumber
\end{align}

The gradients of $q_i(\tau,x)$ for the heat equation with $\sigma_i = \sigma= const$ have been derived first in \citep{LiptonKau2020-2,LiptonKaush2020}, and later in \citep{ItkinMuraveyDB} by using a different method\footnote{These results can be naturally  generalized for the case $\sigma = \sigma(x)$.}. The result read
\begin{align} \label{solUgr}
\fp{q_i(\tau,x)}{x} &\Bigg|_{x = \yi(\tau)} = - \frac{\Psi_i(\tau)}{2 \sigma^3} \left(  \frac{1}{\sqrt{\pi \tau}} + \frac{\yi'(\tau)} {\sigma} \right)   + \int_0^\tau   \frac{\Psi_i(k) e^{-\frac{(\yi(\tau)- \yi(k))^2}{4\sigma^2(\tau - k)}} - \Psi_i(\tau) }{4\sigma^3 \sqrt{\pi (\tau -k)^3}} dk \\
& \hspace{-0.3in} - \int_0^\tau \Psi_i(k) \frac{(\yi(\tau) - \yi(k))^2 e^{-\frac{(\yi(\tau)- \yi(k))^2}{4\sigma^2(\tau - k)}} }{8\sigma^5 \sqrt{\pi (\tau -k)^5}}  dk - \int_0^\tau \Phi_i(k) \cp{G(x,\xi,\sigma^2(\tau-k))}{\xi}{x} \Bigg|_{\substack[b]{\xi = \yii(k)\\x = \yi(\tau)}} dk, \qquad \tau = \tau_i, \nonumber \\
\fp{q_i(\tau,x)}{x}& \Bigg|_{x = \yii(\tau)} = - \int_0^\tau \Psi_i(k) \cp{G(x,\xi,\sigma^2(\tau-k))}{\xi}{x}  \Bigg|_{\substack[b]{\xi = \yi(k)\\x = \yii(\tau)}}  dk - \frac{\Phi_i(\tau)}{2 \sigma^3} \left(\frac{1}{\sqrt{\pi \tau}}  - \frac{\yii'(\tau)}{\sigma} \right)   \nonumber \\
 &\hspace{-0.3in}+ \int_0^\tau   \frac{\Phi_i(k) e^{-\frac{(\yii(\tau)- \yii(k))^2}{4\sigma^2(\tau - k)}} - \Phi_i(\tau) }{4\sigma^3 \sqrt{\pi (\tau -k)^3}} dk  - \int_0^\tau \Phi_i(k) \frac{(\yii(\tau) - \yii(k))^2 e^{-\frac{(\yii(\tau)- \yii(k))^2}{4\sigma^2(\tau - k)}} }{8\sigma^5 \sqrt{\pi (\tau -k)^5}}  dk, \quad
 \tau = \tau_{i+1}, \nonumber
\end{align}
\noindent where $G(x,\xi,\tau)$ is given in \eqref{green}.

\section{Solving the ML heat equation via the GIT method} \label{GIT}

\subsection{Background}\label{Background}

An alternative method to construct the ML problem solution is generalized integral transform (GIT). The GIT method is used in physics, \citep{kartashov1999, kartashov2001}, but was unknown in finance until its first use in \citep{CarrItkin2020jd}. The previously known solution to the heat equation, using the GIT method, was obtained only for the domain $S \in [0,y(t)]$.  For other domains, the solution was unknown even in physics. \citep{ItkinMuravey2020r} were the first to construct the GIT solution for the domain $S \in [y(t), \infty)$. The latter technique was extended further for the CIR and CEV models, \citep{CarrItkinMuravey2020}, the Black-Karasinski model, \citep{ItkinLiptonMuravey}, and finally for double barrier options in \citep{ItkinMuraveyDB}. The latter problem deals with the spatial domain determined by two moving in time boundaries, and boundary conditions, which are arbitrary functions of time.

The GIT and HP methods are similar but have an essential difference. In the HP method, the solution is represented in the form of heat potential with the unknown potential density function $\Psi(\tau)$ which solves the corresponding Volterra equation of the second kind, see Section~\ref{solveMHE}.  In the GIT method, similar to, e.g., the Fourier method, we start with applying some integral transform to the PDE under consideration. The transform has to be such that the transformed equation with $x \mapsto p$ is solvable analytically in time. The second step is to construct an inverse transform, which could be computed analytically using the complex analysis. If this is possible, then the solution can be represented as an explicit integral of some kernel multiplied by the unknown function $\Omega(\tau)$. Hence, this looks pretty similar to the HP method. However, the function $\Omega(\tau)$ has now a transparent meaning - this is the gradient of the solution at the moving boundary. It turns out that this gradient also solves a Volterra equation of the second kind. As mentioned, explicit construction of such forward and inverse transforms is performed in \citep{CarrItkin2020jd, ItkinMuravey2020r, CarrItkinMuravey2020, ItkinLiptonMuravey, ItkinMuraveyDB} for various models and spatial domains. Also, the authors show that the performance of both methods is the same. Both HP and GIT methods are faster than the finite-difference approach and provide higher accuracy.

As mentioned in \citep{ItkinMuraveyDB}, it is not unreasonable to ask why we need two methods - the HP and GIT, which are used to solve the same problem and demonstrate the same performance. The answer is interesting. As shown in \citep{CarrItkinMuravey2020}, the GIT method produces very accurate results at high strikes and maturities (i.e., when the option price is relatively small), in contrast to the HP method, which struggles under these circumstances. One can verify this fact by looking at the exponents under the GIT solution integral {\it proportional} to the time $\tau$. Contrary, when the price is higher (short maturities, low strikes), the GIT method is slightly less accurate than the HP method since the exponents in the HP solution integral are {\it inversely proportional} to $\tau$.

Thus, the GIT and HP methods complement each other for the CIR and CEV models. For other models reducible to the heat equation, the same conclusion holds; see \citep{ItkinMuravey2020r}. This statement is true because GIT integrals contain the difference of two exponents, which becomes small at large $\tau$. On the contrary, the HP exponent tends to one at large $\tau$. Therefore, the convergence properties of the two methods are different at large $\tau$, so they complement each other.

This situation is well known for the heat equation with constant coefficients, \citep{Lipton2001}. There exist two representations of the solution: one - obtained by using the method of images, and the other one - by the Fourier series. Although both solutions are equal when considered as infinite series, their convergence properties are different.

\subsection{Solution of the heat equation}

To apply the GIT method to the solution of \eqref{L_Cauchy_problem1}, we can use the results obtained in \citep{ItkinMuravey2020r}. There it is assumed that $\LL_i$ is the heat equation operator
\begin{equation} \label{LLi_def_GIT}
\LL_i =  - \fp{}{\tau} + \sop{}{x},
\end{equation}
\noindent and then the solution of \eqref{L_Cauchy_problem1} can be represented in the form
\begin{align}  \label{u_final2}
u_i(\tau, x) &= \sum_{n=-\infty}^{\infty} \Bigg \{ \int_{\yi(0)}^{\yii(0)} u_i(0, \xi) \Upsilon_{n,i}(x, \tau  \,|\, \xi, 0) d\xi +\int_0^\tau \left[\Omega_i(s) + \chi_i^{+}(s) \yii'(s) \right]\Upsilon_{n,i} (x, \tau | \yii(s), s)ds, \nonumber \\
&\qquad +\int_0^\tau \left[\Theta_i(s)  - \chi_i^{-}(s) \yi'(s) \right] \Upsilon_{n,i}(x, \tau \,|\, \yi(s), s)ds \\
&\qquad + \int_0^\tau  \chi_i^-(s) \Lambda_{n,i} (x, \tau  \,|\,\yi(s), s) - \chi_i^+(s) \Lambda_{n,i}(x, \tau  \,|\, \yii(s), s) ds \Bigg\}, \nonumber \\
\Upsilon_{n,i}&(x, \tau \,|\, \xi, s) = \frac{1}{2\sqrt{\pi (\tau - s)}}\left[e^{-\frac{(2n l_i(\tau)  +x - \xi)^2}{4 (\tau - s)}} - e^{-\frac{(2n l_i(\tau)  + x +  \xi - 2 \yi(\tau))^2}{4 (\tau - s)}} \right],  \nonumber \\
\Lambda_{n,i}&(x, \tau \,|\, \xi, s)  = \frac{x - \xi + 2n l_i(\tau)}{4 \sqrt{\pi (\tau -s)^3}} e^{-\frac{(2n l_i(\tau)  + x  - \xi)^2}{4 (\tau - s)}} + \frac{x + \xi - 2 \yi(\tau) + 2n l_i(\tau) }{4 \sqrt{\pi (\tau -s)^3}} e^{-\frac{(2n l_i(\tau)  + x +  \xi - 2 \yi(\tau))^2}{4 (\tau - s)}}. \nonumber
\end{align}

Here $\chi_i^-(\tau), \chi_i^+(\tau)$ are the boundary conditions at the left and right boundaries of the $i$-th interval, and
\begin{align} \label{defGrad}
l_i(\tau) &= \yii(\tau) - \yi(\tau), \quad \tau = \tau_i, \\
\Omega_i(\tau) &= -\fp{u_i(\tau, x)}{x} \Bigg|_{x = \yi(\tau)} \qquad \Theta_i(\tau) = \fp{u_i(\tau, x)}{x} \Bigg|_{x = \yii(\tau)}. \nonumber
\end{align}

The functions $\Omega(\tau), \Theta(\tau)$ for the heat equation in \eqref{LLi_def_GIT} can be found by solving a system of the Volterra equations of the second kind. As applied to our problem for the $i$-th interval with $i \in [1,N]$,  it reads, \citep{ItkinMuraveyDB}
\begin{align}  \label{VolterraTheta}
-\Omega_i(\tau) &= \int_{\yi(0)}^{\yii(0)} u(0, \xi) \upsilon_i^-(\tau \, |\, \xi, 0) d\xi \\
&-\frac{\chi_i^-(\tau)}{\sqrt{\pi \tau}} + \int_0^\tau \frac{ \chi_i^-(s) - \chi_i^-(\tau)}{2 \sqrt{\pi (\tau - s)^3}} ds
+   \int_0^\tau  \left[ \chi_i^-(s) d \left(\eta^-_i(\tau \,|\, \yi(s), s)\right)  - \chi_i^+(s) d\left(\eta^-_i(\tau \,|\, \yii(s), s) \right) \right] \nonumber \\
&-\int_0^\tau \Omega_i(s) \frac{\yi(\tau) - \yi(s)}{2\sqrt{\pi (\tau -s)^3}} e^{- \frac{(\yi(\tau) - \yi(s))^2}{4(\tau -s)}} ds
+\int_0^\tau \left[\Theta_i(s) \upsilon_i^-(\tau \,|\,\yii(s), s) + \Omega_i(s)\upsilon^-_{0,i}(\tau \, |\, s)\right] ds \nonumber \\
\Theta_i(\tau) &= \int_{\yi(0)}^{\yii(0)} u(0, \xi) \upsilon_i^+(\tau \, |\, \xi, 0) d\xi \nonumber \\
&+\frac{\chi_i^+(\tau)}{\sqrt{\pi \tau}} - \int_0^\tau \frac{\chi_i^+(s) - \chi_i^+(\tau)}{2 \sqrt{\pi (\tau - s)^3}} ds
+   \int_0^\tau  \left[ \chi_i^-(s) d \left(\eta_i^+(\tau \,|\, \yi(s), s)\right)  - \chi_i^+(s) d\left(\eta_i^+(\tau \,|\, \yii(s), s)\right) \right] \nonumber \\
&-\int_0^\tau \Theta_i(s) \frac{\yii(\tau) - \yii(s)}{2\sqrt{\pi (\tau -s)^3}} e^{- \frac{(\yii(\tau) - \yii(s))^2}{4(\tau -s)}} ds
+\int_0^\tau \left[\Theta_i(s) \upsilon^+_{0,i}(\tau \,|\,s) + \Omega_i(s)\upsilon_i^+(\tau \, |\, \yi(s), s)\right] ds. \nonumber
\end{align}

Here the following notation is used
\begin{align} \label{eta_def}
\eta_i^-(\tau \,|\, \xi, s) &= -\frac{\delta_{\xi, \yi(s)}}{\sqrt{\pi (\tau -s)}}+ \frac{1}{\sqrt{\pi (\tau -s)}} \sum_{n = -\infty}^{\infty} e^{-\frac{(\yi(\tau) - \xi + 2n l_i(\tau))^2}{4(\tau -s)}}, \\
\eta_i^+(\tau \,|\, \xi, s) &= -\frac{\delta_{\xi,\yii(s)}}{\sqrt{\pi (\tau -s)}}+ \frac{1}{\sqrt{\pi (\tau -s)}} \sum_{n = -\infty}^{\infty} e^{-\frac{(\yi(\tau) - \xi + (2n +1) l_i(\tau))^2}{4(\tau -s)}}, \nonumber \\
\upsilon_i^-(\tau \,|\, \xi, s) &= - \sum_{n=-\infty}^\infty \frac{\yi(\tau) - \xi + 2 n l_i(\tau)}{2 \sqrt{\pi (\tau - s)^3}} e^{-\frac{(\yi(\tau) - \xi + 2 n l_i(\tau))^2}{4(\tau - s)}}, \nonumber \\
\upsilon_i^+(\tau \,|\, \xi, s) &= - \sum_{n=-\infty}^\infty \frac{\yi(\tau) - \xi + (2 n  + 1) l_i(\tau)}{2 \sqrt{\pi (\tau - s)^3}} e^{-\frac{(\yi(\tau) - \xi + (2 n + 1) l_i(\tau))^2}{4(\tau - s)}}, \nonumber \\
\upsilon^-_{0,i}(\tau \,|\, s) &= \upsilon^-_{i}(\tau | \yi(s), s) +
\frac{\yi(\tau) - \yi(s)}{2 \sqrt{\pi (\tau - s)^3}} e^{-\frac{(\yi(\tau) - \yi(s))^2}{4(\tau - s)}}, \nonumber \\
\upsilon^+_{0,i}(\tau \,|\, s) &= \upsilon^+_{i}(\tau | \yii(s), s) + \frac{\yii(\tau) - \yii(s)}{2 \sqrt{\pi (\tau - s)^3}} e^{-\frac{(\yii(\tau) - \yii(s))^2}{4(\tau - s)}}, \nonumber
\end{align}
\noindent where $\delta_{\xi, x}$ is the Kronecker symbol. It is worth emphasizing that all summands in \eqref{VolterraTheta} are regular. The integrands in the first two lines have weak (integrable) singularities, while other summands are regular.

At the boundaries of the domain where the solution of our problem is defined, we have
\begin{equation} \label{bcChi}
\chi_1^- = \chi^-,  \qquad \chi_N^- = \chi^+,
\end{equation}
\noindent where $N$ is the number of intervals.

In \citep{ItkinMuraveyDB}, an alternative system of the Volterra equations of the second kind is also obtained, which has the form of \eqref{VolterraTheta}, but with a different definition of the coefficients. We have
\begin{align} \label{eta_def_Theta}
\eta_i^-(\tau \,|\, \xi, s) &=-\frac{\delta_{\xi,\yi(s)}}{\sqrt{\pi (\tau -s)}} + \frac{1}{l_i(\tau)} \theta_3\left( \phi_i(\xi, \yi(\tau)), \omega_i \right), \\
\eta_i^+(\tau \,|\, \xi, s) &=-\frac{\delta_{\xi,\yii(s)}}{\sqrt{\pi (\tau -s)}} + \frac{1}{l(\tau)} \theta_3\left( \phi_i(\xi + l_i(\tau), \yi(\tau)), \omega_i \right),  \nonumber \\
\upsilon_i^-(\tau \,|\, \xi, s) &=-\frac{\pi }{2 l_i^2(\tau)}  \theta'_3\left( \phi_i(\xi, \yi(\tau)), \omega_i \right), \nonumber \\
\upsilon_i^+(\tau \,|\, \xi, s) &= -\frac{\pi }{2 l_i^2(\tau)}\theta'_3\left( \phi_i(\xi + l_i(\tau), \yi(\tau)), \omega_i \right) . \nonumber
\end{align}
Here $\theta_3(z,\omega)$ is the Jacobi theta function of the third kind, \citep{mumford1983tata}, which is defined as follows:
\begin{equation} \label{theta_def}
\theta_3(z,\omega) = 1 + 2\sum_{n = 1}^{\infty} \omega^{n^2} \cos \left(2 n z \right),
\end{equation}
Also, in \eqref{eta_def_Theta} the following notation is used
\begin{align} \label{theta_integrals}
\omega_i &= e^{-\frac{\pi^2 (\tau - s)}{l_i^2(\tau)}}, \qquad \phi_i(x,\xi) = \frac{\pi (x - \xi)}{2 l_i(\tau)}, \\
\fp{\theta_3(z,\omega)}{z} = \theta_3'(z,\omega) &= -4 \sum_{n = 1} ^{\infty} n \omega^{n^2} \sin \left( 2 n z\right). \nonumber
\end{align}

Formulas \eqref{eta_def_Theta} and \eqref{eta_def} are complementary. Since the exponents in the definition of the theta functions in \eqref{eta_def_Theta} are proportional to the difference $\tau - s$, the Fourier series \eqref{eta_def_Theta} converge fast if $\tau - s$ is large. Contrary, the exponents in \eqref{eta_def} are inversely proportional to $\tau - s$. Therefore, the series \eqref{eta_def} converge fast if $\tau - s$ is small.

\subsection{Solution of \eqref{LLi_defND} when $\sigma$ is piecewise constant} \label{sec:sigmax}

Here we assume that $\sigma_i(x) = \sigma_i, \ i=1,\ldots,N$, i.e. the volatility is a piecewise constant function of $x$.
For instance, this is true for the problem described in Section~\ref{Dupire}. As shown there, the pricing PDE   can be transformed to \eqref{LLi_def} instead of \eqref{heatHW}. According to the transformation in \eqref{trW} the clock will run differently at each ML interval, which is inconvenient. Therefore, instead of a change of temporal variable, below we use a transformation of the spatial variable $x$. This transformation allows using the same time at each ML interval. To achieve this, we change the definition of $\tau(t)$ in \eqref{trW} to
\begin{equation}
	\tau = \frac{1}{2} \int_0^T  e^{-2 \int_0^s [r(s) - q(s)] \, dk} \, ds,
\end{equation}
\noindent where $r(t), q(t)$ for the problem considered in Section~\ref{Dupire} are the deterministic interest rate and continuous dividends. This converts the problem in Section~\ref{Dupire} and the PDE \eqref{dupFin2} to
\begin{align} \label{DupireProblem}
	\fp{U(\tau,x)}{\tau} &= \sigma^2(x) \sop{U(\tau,x)}{x}, \\
	U(0,x) &= U_0(x) = (x - S)^+, \nonumber \\
	U(\tau,0) &= 0, \qquad U(\tau,x)\Big|_{x \to \infty}  =
	x - e^{-\int_0^T (r(s) - q(s)) ds} S. \nonumber
\end{align}
And, according to \eqref{approx}
\begin{equation}\label{key}
	\sigma^2(T,K) = v_i(T), \quad K \in [K_i, K_{i+1}].
\end{equation}
The boundary and initial conditions in \eqref{DupireProblem} are the direct translation of those conditions in \eqref{tcHW} and \eqref{bc}.

Again, as shown in Appendix~\ref{divA}, the problem in \eqref{DupireProblem} can be transformed to
\begin{align} \label{DupireProblem1}
	\fp{U(\tau,\hx)}{\tau} &= \fp{}{\hx}\left(\Xi^2(\hx) \fp{U(\tau,\hx)}{\hx}\right), \qquad \hx = \hx(x), \\
	U(0,\hx) &= U_0(\hx) = (x(\hx) - S)^+, \nonumber \\
	U(\tau,\hx(0)) &= 0, \qquad U(\tau,x(\hx))\Big|_{x(\hx) \to \infty}  =
	x(\hx)  - e^{-\int_0^T (r(s) - q(s)) ds} S. \nonumber
\end{align}

 Also, since based on \eqref{trW}
\begin{equation*}
U(\tau,x(\hx)) = P(T,K) e^{-\int_0^T q(s) \, ds},
\end{equation*}
\noindent the continuity of the solution and its flux at all internal boundaries can be expressed as
\begin{align} \label{cont2}
\chi^+_i(\tau) &= \chi^-_{i+1}(\tau). \\
  \Xi_{i+1}^2 \Omega_{i+1}(\tau)  & = - \Xi^2_{i}\Theta_{i}(\tau). \nonumber
\end{align}

To use the results of the previous Section, we proceed by applying the following transformation to \eqref{u_final2}
\begin{equation} \label{traSigma}
\bx = \Xi_i \hx,\qquad \by(\tau) = \Xi_i y(\tau), \qquad \bxi = \Xi_i \xi,
\qquad \qquad \bl(\tau) \equiv \Xi_i l(\tau).
\end{equation}
Note, that the last equality in \eqref{traSigma} is actually the definition of $\bl(\tau)$. Another complication which comes due to this transformation is that in new variables $\bar{x}$ the layers stop to be continuous. In other words, the upper boundary of the $i$-th layer $\byi^+(\tau)$ and the lower boundary of the $(i+1)$-th layer $\byii^-(\tau)$ are now not equal. Therefore, in what follows to avoid any confusion we will explicitly use this notation, i.e. the left and right boundaries of the $i$-th layer are denoted as $\yi^-(\tau)$ and $\yii^-(\tau)$.

Also, per these transformations, we have
\begin{align}
u_i(0,\bxi) = u_i(0,\Xi_i, \xi), \qquad u_i(0,\Xi_i, \xi) d\xi &= \frac 1\Xi_i u_i(0,\bxi) d\bxi, \\
\int_{\yi(0)}^{\yii(0)} u(0, \xi) \upsilon_i^-(\tau \, |\, \xi, 0) d\xi &=
\frac{1}{\Xi_i}\int_{\byi(0)}^{\byii(0)} u(0, \bxi) \bar{\upsilon}_i^-(\tau \, |\, \bxi, 0) d\bxi, \nonumber \\
\bar{\upsilon}_i^-(\tau \, |\, \bxi, 0) &= - \sum_{n=-\infty}^\infty \frac{\byi^-(\tau) - \bxi + 2 n \bl_i(\tau)}{2 \Xi_i \sqrt{\pi (\tau - s)^3}} e^{-\frac{(\byi^-(\tau) - \bxi + 2 n \bl_i(\tau))^2}{4 \Xi_i^2(\tau - s)}}. \nonumber
\end{align}

It is easy to check that this transformation leaves $\phi_i(\hx,\xi)) $ and $\partial_\hx (\phi_i(\hx,\xi))$ invariant,  but with the new definition
\begin{equation*}
\bar{\omega}_i = e^{-\frac{\pi^2 \Xi_i^2 (\tau - s)}{\bl_i^2(\tau)}}.
\end{equation*}
Finally, let us redefine the partial derivatives
\begin{equation} \label{gradNew}
\bar{\Omega}_i = -\fp{u(\tau, \bx)}{\bx} \Bigg|_{\bx = \byi^-(\tau)}, \qquad \bar{\Theta}_i = \fp{u(\tau, \bx)}{\bx} \Bigg|_{\bx = \byi^+(\tau)}
\end{equation}
\noindent instead of their definitions in \eqref{defGrad}, i.e. $\Omega_i = \Xi_i \bar{\Omega}_i, \ \Theta_i = \Xi_i \bar{\Theta}_i$. Then the continuity conditions in \eqref{cont2} change to
\begin{align} \label{cont21}
\bchi^+_i(\tau) &= \bchi^-_{i+1}(\tau), \\
  \Xi^3_{i+1} \bOmega_{i+1}(\tau)  & = - \Xi^3_{i}\bTheta_{i}(\tau). \nonumber
\end{align}

To simplify notation, we omit bars from all new variables assuming this doesn't bring any confusion. Then \eqref{u_final2} transforms to
\begin{align}  \label{u_final3}
u_i(\tau, x) &= \sum_{n=-\infty}^{\infty} \Bigg \{ \frac{1}{\Xi_i}\int_{\yi(0)}^{\yii(0)} u_i(0, \xi) \Upsilon_{n,i}(x, \tau  \,|\, \xi, 0) d\xi \\
&\qquad +\int_0^\tau \left[\frac{1}{\Xi_i}\Omega_i(s) + \frac{1}{\Xi_{i+1}}\chi_i^{+}(s) \yi^{+'}(s) \right]\Upsilon_{n,i} (x, \tau | \yi^+(s), s)ds, \nonumber \\
&\qquad +\frac{1}{\Xi_i}\int_0^\tau \left[\Theta_i(s)  - \chi_i^{-}(s) \yi^{-'}(s) \right] \Upsilon_{n,i}(x, \tau \,|\, \yi^-(s), s)ds \nonumber \\
&\qquad + \int_0^\tau  \chi_i^-(s) \Lambda_{n,i} (x, \tau  \,|\,\yi^-(s), s) - \chi_i^+(s) \Lambda_{n,i}(x, \tau  \,|\, \yi^+(s), s) ds \Bigg\}, \nonumber \\
\Upsilon_{n,i}(x, \tau \,|\, \xi, s) &= \frac{1}{2\sqrt{\pi (\tau - s)}} \left[e^{-\frac{(2n l_i(\tau)  +x - \xi)^2}{4\Xi^2_i (\tau - s)}} - e^{-\frac{(2n l_i(\tau)  + x +  \xi - 2 \yi^-(\tau))^2}{4 \Xi^2_i (\tau - s)}} \right],  \nonumber \\
\Lambda_{n,i}(x, \tau \,|\, \xi, s)  &= \frac{x - \xi + 2n l_i(\tau)}{4 \Xi_i\sqrt{\pi (\tau -s)^3}} e^{-\frac{(2n l_i(\tau)  + x  - \xi)^2}{4 \Xi^2_i(\tau - s)}} + \frac{x + \xi - 2 \yi^-(\tau) + 2n l_i(\tau) }{4 \Xi_i\sqrt{\pi (\tau -s)^3}} e^{-\frac{(2n l_i(\tau)  + x +  \xi - 2 \yi^-(\tau))^2}{4 \Xi^2_i(\tau - s)}}. \nonumber
\end{align}
By analogy, the modified Volterra equations can be obtained from \eqref{VolterraTheta}.

In \eqref{VolterraTheta} the unknown variables are $[\chi^-_1(\tau), \chi^+_1(\tau), \Omega_1(\tau), \Theta_1(\tau), \ldots, \chi^-_N(\tau), \chi^+_N(\tau), \Omega_N(\tau), \Theta_N(\tau)]$, so that there are $4 N$ unknowns in total. The boundary conditions in \eqref{bcChi} and the continuity conditions in \eqref{cont2} reduce the number of unknown variables to $2N-2$, because $\chi^+_i(\tau), \Theta_{i}(\tau)$ can be expressed via  $\chi^-_i(\tau), \Omega_{i}(\tau)$ and substituted into \eqref{VolterraTheta}. Thus, the GIT method provides a significant simplification of the system of Volterra equations as compared with the HP method.

\section{Application to finance} \label{secApp}

In this section, we consider several models that are frequently used in mathematical finance. We provide a short description of each model and demonstrate how to reduce the corresponding pricing problem to the form suitable for solving it by the ML method.

\subsection{One-factor short-rate models}

 As the first example, we consider one-factor short interest rate (IR) models. Although these models were developed a long time ago, they are still essential and widely used by practitioners. While one can price zero-coupon bonds (ZCB) and European options on the ZCB and swaptions for many of them analytically, this is not true for exotic options. For instance, pricing of barrier options when the barriers are time-dependent and could pay time-dependent rebates has to be done numerically. The same is true for American options.

 However, as mentioned in the Introduction, one can find the solution to these problems semi-analytically using the HP and GIT methods for some one-factor models, including the time-dependent OU (Vasicek) model in
 \citep{CarrItkin2020jd,LiptonKau2020-2,LiptonKaush2020}, for the Hull-White model in \citep{ItkinMuravey2020r}, for the CEV and CIR models in  \citep{CarrItkinMuravey2020}, and then in a general form for any model that can be reduced to the heat equation - in \citep{ItkinMuraveyDB}. In other words, solving these problems doesn't require the ML method. Therefore, below we consider some other models for which the barrier pricing problems cannot be directly solved by the HP or GIT methods but can be solved by using the ML method.

\subsubsection{Pricing zero-coupon bonds and barrier options for the Black-Karasinski and similar models} \label{genBK}

The Black-Karasinski (BK) model was introduced in \citep{BK1991}, see also \citep{BM2006} for a more detailed discussion. The BK is a one-factor short interest rate model of the form
\begin{align} \label{BK}
d z_t &= k(t)[\theta(t) - z_t] dt + \sigma(t)dW_t, \qquad r \in \mathbb{R}, \ t \ge 0, \\
r_t &= s(t) + R e^{z_t}, \qquad r(t=0) = r_0. \nonumber
\end{align}
Here $\kappa(t) > 0$ is the constant speed of mean-reversion, $\theta(t)$ is the mean-reversion level, $\sigma(t)$ is the volatility, $R$ is some constant  with the same dimensionality as $r_t$, eg., it can be 1/(1 year). This model is similar to the Hull-White model but preserves the positivity of $r_t$ by exponentiating the Ornstein-Uhlenbeck (OU) random variable $z_t$. Because of that, usually, practitioners add a deterministic function (shift) $s(t)$ to the definition of $r_t$ to address possible negative rates and be more flexible when calibrating the term-structure of the interest rates.

By \Ito lemma the short rate $\bar{r}_t = (r_t - s(t))/R$ in the BK model solves the following stochastic differential equation (SDE)
\begin{equation} \label{BKr}
d \bar{r}_t = [k\theta(t) + \frac{1}{2}\sigma(t)^2 - k \log \bar{r}_t] \bar{r}_t dt + \sigma(t) \bar{r}_t dW_t.
\end{equation}
\noindent This SDE can be explicitly integrated. Let $0 \le s \le t \le T$, Then $r_t$ can be represented as, \citep{BM2006}
\begin{equation} \label{BKsol}
\bar{r}_t = \exp\left[
e^{-k(t-s)} \log \bar{r}_s + k \int_s^t e^{-k(t-u)} \theta(u) du + \int_s^t \sigma(u) e^{-k(t-u)} dW(u)\right],
\end{equation}
\noindent and thus, conditionally on filtration $\mathcal{F}_s$ is lognormally distributed and always positive.

However, in the BK model, the price $P(t,T)$ of a (ZCB) with the maturity $T$ is not known in closed form since this model is not affine. Multiple good approximations have been developed in the literature using asymptotic expansions of various flavors, see, e.g., \citep{AntonovSpector2011,Capriotti2014,Horvath2017}, and also survey in \citep{Turfus2020}.

Despite this lack of tractability, the BK model is widely used by practitioners for modeling interest rates and credit and is also known in commodities as the  Schwartz one-factor model. The BK model is attractive because it is relatively simple, guarantees non-negativity of the prices (which could be a bad feature in the current environment). It could also be calibrated to the given term-structure of interest rates and the prices or implied volatilities of caps, floors, or European swaptions since the mean-reversion level and volatility are functions of time. However, for exotic options, e.g., highly liquid barrier options, these prices are not known yet in closed form. Therefore, various numerical methods are used to obtain them.

Here we describe how one can reduce the pricing problem for the ZCB to the ML heat equation. Since this problem is defined at a semi-infinite domain,  the corresponding ML heat equation is also defined at a semi-infinite interval. Thus, the number of layers could be infinite. Therefore,  truncation of the semi-infinite interval to a finite is needed. Of course, the impact of the remainder should be assessed appropriately.

Along the BK model lines, consider a model where the dynamics of the underlying stochastic variable $z_t$ is the OU process defined in \eqref{BK}. We assume that the interest rate $r_t$ is some deterministic function of $z_t$
\begin{equation}
r_t = s(t) + f(t,z_t), \qquad z_0 = 0.
\end{equation}
In particular, according to \eqref{BK} for the BK model we have $f(t,z_t) = R e^{z_t}$, and so $R = r_0 - s_0$.

In terms of $z$, the corresponding PDE for the ZCB price $F(t,r)$ in \eqref{PDEBK} and for the option price $C(t,r)$ in \eqref{PDEP} reads
\begin{align} \label{VPDE}
0 &= \fp{V}{t} + \dfrac{1}{2}\sigma^2(t) \sop{V}{z} + \kappa(t) [\theta(t) - z] \fp{V}{z} - [s(t) + f(t,z)] V,
\end{align}
\noindent where $V = V(t,z)$ is either $F(t,z)$ or $C(t,z)$. This equation should be solved subject to the terminal and boundary conditions. For the ZCB price they are given in \eqref{termZCB} and  \eqref{bcZCB}, and for the Down-and-out barrier Call option price - in  \eqref{tc0BK} and in \eqref{bcBar}, \eqref{bc01}. Note, that solving \eqref{VPDE} for $F(t,z)$ assumes that $z \in \left(f^{-1}(t,-\infty), f^{-1}(t,\infty)\right)$, while for $C(t,z)$ the domain of definition is $z \in [L_z(t), f^{-1}(t,\infty))$., where $L_z(t) = f^{-1}(t,L(t))$, and $f^{-1}(t,r)$ is the inverse function.

To apply the ML method to \eqref{VPDE}, for instance, when solving the barrier option pricing problem, we truncate the interval $[L(t),\infty)$ from above at $z = z_{\max}$ to make it $[L_z(t),z_{\max}]$. The reason this is possible lies in the fact that when $z$ increases, the ZCB price tends to zero based on the boundary condition. Therefore, the Call option price in this limit vanishes as well. Thus, the contribution of the region $[z_{\max},f^{-1}(t,\infty))$ to the Call option price becomes negligible
\footnote{For some choices of the functions $f(t,z)$ the value $f^{-1}(t,\infty)$ could be finite which eliminates the need for truncation.}.

Now we split the interval $[L_z(t),z_{\max}]$ into $N > 0$ sub-intervals, and at every interval $[z_i,z_{i+1}], \ i=1,\ldots,N$ assume that $f(t,z)  = a_i(t) + b_i(t) z)$. Accordingly, at every interval $i, \ i=1,\ldots,N$ the PDE \eqref{VPDE} takes the form
\begin{align} \label{VPDE1}
0 &= \fp{V}{t} + \dfrac{1}{2}\sigma^2(t) \sop{V}{z} + \kappa(t) [\theta(t) - z] \fp{V}{z} - [s(t) + a_i(t) + b_i(t) z] V.
\end{align}
This PDE can be transformed to the heat equation
\begin{align} \label{heat1}
\fp{U}{\tau} &= \sop{U}{x},
\end{align}
\noindent by the change of variables, \citep{Polyanin2002,ItkinMuravey2020r,LiptonKaush2020}
\begin{equation} \label{transH}
V(t,z) = \exp[\alpha_i(t) z + \beta_i(t)] U(\tau,x), \qquad \tau = \phi(t), \qquad x = z \psi(t) + \varrho(t),
\end{equation}
\noindent where
\begin{alignat}{2} \label{coef}
\psi(t) &= C_1 \exp\left( \int_S^t \kappa(q) dq \right), \qquad &
\phi(t) &= \frac{1}{2} \int_t^S  \sigma^2(q) \psi^2(q) d q + C_2, \\
\alpha_i(t) &= \psi(t) \int_S^t \frac{b_i(q)}{\psi(q)} dq + C_3 \psi(t), \qquad &
\varrho_i(t) &= -\int_S^t  \left[ \kappa(q) \theta(q) + \sigma^2(q) \alpha_i(q) \right] \psi(q) dq + C_5,  \nonumber
\end{alignat}
\begin{equation*}
\beta_i(t) = - \frac{1}{2} \int_S^t \alpha_i(q) \left[2 \kappa(q) \theta(q) + \sigma^2(q) \alpha_i(q) \right] dq
+ \int_S^t  [s(q) + a_i(q)]dq + C_4,
\end{equation*}
\noindent where $C_1,\ldots,C_5$ are some constants. In our case we can choose $C_1 = 1$, $C_2 = C_3 = C_4 = C_5  = 0$.

One of the advantages of such an approach is that the new time $\tau$ doesn't depend on the specific interval $i$, i.e. the time $\tau$ runs in sync for all intervals $[z_i,z_{i+1}], \ i=1,\ldots,N$.

Thus, the main trick here is in using the approximation $f(t,z)  = a_i(t) + b_i(t) z)$. This approximation provides the second order of accuracy in the length of the interval (similar to the finite-difference method of the second order), and allows reduction of the PDE at each interval to the heat equation (while the original PDE doesn't hold this property).

At the end of this Section, note that \eqref{VPDE1}, if used for pricing ZCB, doesn't need the boundary condition at the left boundary $z \to -\infty$, as this is discussed in \citep{ItkinMuravey2020r} with a reference to Fichera theory, \citep{OleinikRadkevich73}. However, the price of the ZCB at some fixed left boundary $z_{\min}$, i.e. $V(t,z_{\min})$ can be found having in mind that the transformed PDE in \eqref{VPDE1} is affine, which yields
\begin{equation} \label{affSol}
V(z,t,S) = A(t,S) e^{ B(t,S) R e^z}.
\end{equation}

With allowance for the terminal condition in \eqref{termZCB}, the solution reads, \citep{ItkinMuravey2020r} \begin{align} \label{ZCBsol}
B(t,S) &= e^{\int_0^t \kappa(m) \, dm} \int_S^t b_i(m) e^{-\int_0^m \kappa(q) \, dq} \, dm, \\
A(t,S) &= \exp\left[\int_S^t \left(a_i(m) + s(m) -\frac{1}{2}  B(m,S) \left( 2 \theta(m) \kappa(m) + B(m,S) \sigma^2(m) \right) \right) dm \right]. \nonumber
\end{align}
It can be seen that $B(t,S) < 0$ if $t < S$. Therefore,  $F(r,t,S) \to 0$ when $z \to \infty$.

\subsubsection{The modified BK (Verhulst) model}

Since the BK model is not fully tractable, in \citep{ItkinLiptonMuravey} we introduced a slightly modified version of the model as follows
\begin{align} \label{BK1}
d z_t &= k(t)[\bar{\theta}(t) - e^{z_t}] dt + \sigma(t)dW_t, \\
r_t  &=  s(t) + R e^{z_t}, \qquad z_0 = 0, \qquad R = r_0 - s(0). \nonumber
\end{align}
It can be seen, that at small $t$ $|z_t| \ll 1$, and so choosing $\bar{\theta}(t) = 1 + \theta(t)$ replicates the BK model in the linear approximation on $z_t$. Similarly, the choice $\bar{\theta}(t) = e^{\theta(t)}$ replicates the BK model at $z_t$ close the mean-reversion level $\theta(t)$. Thus, this model acquires the properties of the BK model while is a bit more  tractable as this will be seen below.

It is worth noting that if by using \Ito lemma we re-write \eqref{BK1} for the stochastic variable $r_t$, the resulting dynamics
can be recognized as the stochastic Verhulst or stochastic logistic model, which are well-known in the population dynamics and epidemiology; see, eg., \citep{Verhulst, bacaer2011, Logistic2015} and references therein. For more information, see \citep{ItkinLiptonMuravey}.

By the \Ito lemma and the Feynman–Kac formula any contingent claim written on the $r_t$ as the underlying (for instance, price $F(\br,t,S)$ of a Zero-coupon bond (ZCB) with maturity $S$) solves the following partial differential equation
\begin{align} \label{PDEBK}
0 &= \fp{F}{t} + \dfrac{1}{2}\sigma^2(t) \br^2 \sop{F}{\br} + \kappa(t) \br [\tth(t) - \br] \fp{F}{\br} - (s(t) + R \br)F, \\
\br_t &= \frac{r_t - s(t)}{r_0 - s(0)} = e^{z_t}, \qquad \tth(t) = \bth(t) + \frac{1}{2}\sigma^2(t). \nonumber
\end{align}
This equation should be solved subject to the terminal condition
\begin{equation} \label{termZCB}
 F(\br,S,S)  = 1,
\end{equation}
\noindent and the boundary condition
\begin{equation} \label{bcZCB}
F(\br,t,S)\Big|_{\br \to \infty} = 0,
\end{equation}
\noindent see, eg., \citep{andersen2010interest}.

In the sequel we will also consider a Down-and-Out barrier Call option written on the ZCB. It is known, \citep{andersen2010interest}, that under a risk-neutral measure the option price $C(t,\br)$ solves the same PDE as in \eqref{PDEBK},
\begin{equation} \label{PDEP}
0 = \fp{C}{t} + \dfrac{1}{2}\sigma^2(t) \br^2 \sop{C}{\br} + \kappa(t) \br [\tth(t) - \br] \fp{C}{\br} - (s(t) + R \br)C.
\end{equation}
The terminal condition at the option maturity $T \le S$ for this PDE reads
\begin{equation} \label{tc0BK}
C(T,\br) = \left(F(\br,T,S) - K\right)^+,
\end{equation}
\noindent where $K$ is the option strike.

By a standard contract, the lower barrier $L_{F}(t)$ (which we assume to be time dependent as well) is set on the ZCB price, and not on the underlying interest rate $\br$. This means that it can be written in the form
\begin{equation} \label{ZCBBar}
C(t,\br) = 0 \quad \mbox{if } F(\br,t,S) = L_F(t).
\end{equation}
This condition can be translated into the $\br$ domain by solving the equation
\begin{equation*}
F(\br,t,S) =  L_F(t),
\end{equation*}
\noindent with respect to $\br$. Denoting the solution of this equation as $L(t)$ we find that \eqref{ZCBBar} in the $\br$ domain reads
\begin{equation} \label{bcBar}
C(t,L(t)) = 0.
\end{equation}
The second boundary can be naturally set at $\br \to \infty$. As at $\br \to \infty$ the ZCB price tends to zero, the Call option price also vanishes in this limit. This yields
\begin{equation} \label{bc01}
 C(t,\br)\Big|_{\br \to \infty}  = 0.
\end{equation}
Accordingly, \eqref{PDEP} has to be solve at $\br \in [L(t), \infty)$.

\subsubsection{Pricing barrier options in the Verhulst model} \label{MBKeq}

As we have already mentioned, in contrast to other similar one-factor models like the time-dependent Ornstein-Uhlenbeck, Hull-White, CIR and CEV models which have been considered in \citep{CarrItkin2020jd, ItkinMuravey2020r, CarrItkinMuravey2020}, the solution of the pricing problem for the BK model is not known in closed form\footnote{Some approaches for doing that are discussed in \citep{ItkinLiptonMuravey}.}. Therefore, we propose an approximation that gives rise to a semi-analytical solution for the barrier Call option price. This approximation is inspired by the ML heat equations which are discussed in Section~\ref{layer}.

Since our problem in \eqref{PDEP} is defined at the semi-infinite domain $\br \in [L(t), \infty)$, using the ML approximation is time-consuming, as we need to split this semi-infinite interval into a fixed number of sub-intervals. Therefore, it is feasible first to make a change of variables
\begin{equation} \label{tr1}
C(t, \br) = V(t,x) e^{\int_0^t s(k) dk}, \qquad x = \frac{a(t)}{\br}, \qquad a(t) = e^{\int_0^t \left( \kappa (m) \tth(m)-\sigma^2 (m)\right) \, dm},
\end{equation}
\noindent so the problem to solve in new variables reads
\begin{align} \label{PDE2}
0 &= \fp{V}{t} + \dfrac{1}{2}\sigma^2(t) x^2 \sop{V}{x} +  a(t) \kappa(t) \fp{V}{x} - R a(t)\frac{V}{x}.
\end{align}
Thus, now our problem is defined at a fixed domain $x \in [0,1/L(t)]$, where the upper boundary is time dependent. Accordingly, in the new variables the Down-and-Out barrier Call option becomes the Up-and-Out barrier Call.

Doing the second change of the dependent variable
\begin{align} \label{tr2}
V(t,x) &= u(t,x) e^{d(t)/ x}, \qquad d(t) = R e^{-\int_0^t \sigma^2(m) dm} \int_0^t e^{\int_0^y \kappa(m) \tth(m) dm} dy
\end{align}
\noindent yields the equation
\begin{align} \label{PDE3}
0 &= \fp{u}{t} + \dfrac{1}{2}\sigma^2(t) x^2 \sop{u}{x} + g(t) \fp{u}{x} - f(t)\frac{u}{x^2}, \\
f(t) &= \frac{1}{2} d(t)\left(2 a(t) \kappa(t) - d(t) \sigma^2(t)\right), \qquad g(t) = a(t) \kappa(t) - d(t) \sigma^2(t). \nonumber
\end{align}
Accordingly, in the new variables the initial and boundary conditions read
\begin{align} \label{tcbc}
u(T,x) &= \exp\left(-\frac{d(T)}{x} - \int_0^T s(k) dk \right) \left(F(x,T,S) - K\right)^+, \\
u(t,0) &= 0, \qquad u\left(t, y(t)\right) = 0, \quad y(t) = a(t)/L(t). \nonumber
\end{align}

The problem in \eqref{PDE3}, \eqref{tcbc} cannot be solved in closed form. Therefore, we proceed by borrowing the idea from the ML approach in physics which is described in Sections~\ref{solveMHE},\ref{GIT}. This approach implies that the interval $x \in [0,y(t)]$ we approximate the function $\zeta(x) = x^2$ by using a piecewise constant approximation. In more detail, we split the interval $[0,y(t)]$ into $N > 0$ sub-intervals, and at every interval $[x_i,x_{i+1}], \ i=1,\ldots,N$ assume that $x^2 \approx \nu_i(t)$\footnote{Since the upper boundary of the whole interval $y(t)$ is the function of time, we need to put $\nu_i = \nu_i(t)$}. For instance, one can choose the middle value of the function $\zeta(x)$ at each sub- , so
\begin{equation*}
\nu_i(t) = y^2(t) \frac{(i+1/2)^2}{N^2}.
\end{equation*}

With allowance for this approximation, at every $i$-th interval \eqref{PDE3} takes the form
\begin{align} \label{PDEap}
0 &= \fp{u}{t} + \dfrac{1}{2}\sigma^2(t) \nu_i(t) \sop{u}{x} + g(t) \fp{u}{x} - \frac{f(t)}{\nu_i(t)} u.
\end{align}
The \eqref{PDEap} can be transformed to the heat equation
\begin{align} \label{heat}
\fp{U}{\tau} &= \sop{U}{\varsigma},
\end{align}
\noindent using the transformation, \citep{Polyanin2002}
\begin{align}
u(t,x) = U(\tau,\varsigma) \exp\left(- \int_0^t \frac{f(t)}{\nu_i(t)} \right), \qquad l = x - \int_0^t g(k) dk, \qquad
\tau = \frac{1}{2} \int_t^T \sigma^2(k) \nu_i(k) dk.
\end{align}

Note, that under this approximation the new time $\tau$ also becomes a function of the interval $i$.

\subsection{Local volatility and Dupire's equation} \label{Dupire}

Calibration of the local volatility model (constructed by using a one-factor Geometric Brownian motion process) to a given set of option prices is a classical problem of mathematical finance. It was considered in multiple papers, and various solutions were proposed; see, e.g., a survey in \citep{ItkinLocalVol,ItkinLipton2017} and references therein.  In particular, in \citep{ItkinLipton2017} an analytical approach to solving the calibration problem is developed. This approach extends the method in \citep{LiptonSepp2011iv} by replacing a piecewise constant local variance construction with a piecewise linear one and allowing non-zero interest rates and dividend yields. This approach remains analytically tractable as it combines the Laplace transform in time with an analytical solution of the resulting spatial equations in terms of Kummer's degenerate hypergeometric functions.

A similar problem could be formulated not just for the Black-Scholes model but also for other models. For instance, in  \citep{ELVG,GLVG} two extensions of the Local Variance Gamma model proposed initially in \citep{CarrNadtochiy2017} were developed. The first new model (ELVG) considers a Gamma time-changed {\it arithmetic} Brownian motion with drift and the local variance to be a piecewise linear function of the strike. The second model (GLVG) is a {\it geometric} version of the ELVG with drift. It also treats various cases by introducing three {\it piecewise linear} models: the local variance as a function of strike, the local variance as a function of log-strike, and the local volatility as a function of strike (so, the local variance is a {\it piecewise quadratic} function of the strike). For all these extensions, the authors derive an ordinary differential equation for the option price, which plays the role of Dupire's equation for the standard local volatility model. Moreover, it can be solved in closed form.

In \citep{ItkinLipton2017,ELVG,GLVG} all models were calibrated to the market quotes term-by-term. Therefore, various types of no-arbitrage interpolation were proposed to guarantee no-arbitrage while keeping the model analytically tractable on the other hand; further details are given in \citep{ItkinLocalVol}.

Two advantages of the semi-analytical approach, which are essential for calibration the model, should be emphasized. First, the option prices can be found analytically in a semi closed form. Here "semi" means that the analytic solution requires an additional inverse transform to be applied to get the final prices; see \citep{ItkinLipton2017}. However, in \citep{ELVG,GLVG} since the Dupire-like equation is an ODE and not a PDE, this step is eliminated. Nevertheless, all these models are calibrated term-by-term.

This idea can be extended by constructing a semi-analytical solution of the ML heat equation, which is analytic in time. Thus, the term-by-term calibration could be eliminated, and the quotes for all strikes and maturities can be used simultaneously. Therefore, this approach allows a further acceleration of the calibration process.

For brevity, let us consider European options, for instance, a Put option on a stock. It is well-known that the price $P(T,K)$  of the option written on the underlying stock price $S_t$ as a function of the option maturity $T$ and strike $K$ solves Dupire's equation, \citep{Dupire:94}
\begin{align} \label{dupFin1}
\fp{P}{T} &= \frac{1}{2} \sigma^2(T,K) \sop{P}{K} - (r-q)K \fp{P}{K} - q(T) P.
\end{align}
\noindent with $\sigma = \sigma(t,S)$ being the local volatility, and $q(t)$ is the dividend yield. This PDE should be solved subject to the initial condition at $T=0$
\begin{equation} \label{tcHW}
P(0,K) = (K - S)^+,
\end{equation}
\noindent and natural boundary conditions for the put option price that read, \citep{Hull2011}
\begin{equation} \label{bc}
\begin{array}{lll}
P(T,K) = 0, & & K \to 0, \\
P(T,K) = D(t,T) (K - S) \approx D(t,T) K, & & K \to \infty,
\end{array}
\end{equation}
\noindent where the discount factor $D(t,T)$ is defined as
\begin{equation} \label{df_01}
D(t,T) = \exp\left(-\int_t^T r(k) dk \right).
\end{equation}

To proceed further, we use the idea in \citep{LiptonSepp2011iv,ItkinLipton2017} and approximate the local variance using some piecewise approximation in the strike space. However, in contrast to \citep{LiptonSepp2011iv,ItkinLipton2017} we make this approximation a function of time. Further, suppose that for each trading maturity $T_j, \ j \in [1,M]$ the market quotes are provided at a set of strikes $K_{i,j}, \ i=1,\ldots,n_j$ where the strikes are assumed to be sorted in the increasing order. Let us construct a finite grid in the strike space $\mathbf{G}(K): \, K \in [\min(K_{i,j}),\max(K_{i,j})], \ j =1,\ldots,M, \ i =1,\ldots,n_j$, by splitting the whole interval $K \in [\min(K_{i,j}),\max(K_{i,j})]$ into $n$ sub-intervals. At every interval $[K_i, K_{i+1}], \ i=1,\ldots,N$, we approximate the local variance function $\sigma^2(T,K)$ by a piecewise constant function in $K$ as follows:
\begin{equation} \label{approx}
\sigma^2(T,K) = v_i(T), \quad K \in [K_i, K_{i+1}].
\end{equation}
This approximation is not continuous, so the local variance $\sigma^2(T,K)$ experience a finite jump at every point $K_i$. However, it is continuous in the maturity $T$.

Accordingly, at every interval $[K_i,K_{i+1}]$ \eqref{dupFin1}, $V(T,K)$ takes the form
\begin{align} \label{dupFin2}
\fp{P_i}{T} &= \frac{1}{2} v_i(T) \sop{P_i}{K} - [r(T) - q(T)] \fp{P_i}{K} - q(T) P_i.
\end{align}
This equation can be transformed to the heat equation
\begin{equation} \label{heatHW}
\fp{U_i}{\tau} = \sop{U_i}{x},
\end{equation}
\noindent by a change of variables
\begin{align} \label{trW}
P_i(T,K) &= U_i(\tau,x) e^{-\int_0^T (q(s) ds}, \quad x =  e^{-\int_0^T (r(s) - q(s) \, ds} K, \quad \tau = \frac{1}{2} \int_0^T  v_i(s) e^{-2 \int_0^s (r(s) - q(s)) \, dk} \, ds.
\end{align}
It is important to note that the new time $\tau$ runs differently at every interval $K \in [K_i, K_{i+1}]$ as it depends on the local variance value $v_i(s)$ at this interval.

\section{Solution of the Volterra equations} \label{solVolt}

An efficient solution of the derived systems of Volterra equations is a problem that requires some attention and extended description. Therefore, it will be published elsewhere. Instead, here we show that a particular choice of the internal boundaries can reduce this problem to a linear system with a tridiagonal matrix allowing the inverse Laplace transform. In this section, we explain this approach in detail.

We start with \eqref{VolterraTheta}. Using the definitions of $\eta^\pm, \upsilon^\pm, \upsilon_0^\pm$ in \eqref{eta_def}, we observe that under all integrals in \eqref{VolterraTheta} the functions $f_i(s), \Theta_i(s), \Omega_i(s)$ are functions of $s$, while functions $\eta^\pm, \upsilon^\pm, \upsilon_0^\pm$ are functions of $t-s$ and $y_{i}(t) - y_i(s)$ and $z_{i}(t) - z_i(s)$. Recall that functions $y_1(t) = \chi^-, \ y_N(t) = \chi^+$ define the external boundaries of the computational domain, while functions $y_i(t), \ i=2,\ldots,N-1$ define the boundaries of the internal layers (the internal boundaries).

Since the internal boundaries are artificial, we can construct them as we wish. For instance, we can use polynomial functions, such as $y_i(s) = a_i s^2 + b_i s + c_i$, where $a_i, b_i, c_i$ are some constants. Then $y_i(t) - y_i(s)$ can also be represented as a certain function $g(t-s)$. Indeed
\begin{equation}
 y_i(t) - y_i(s) = -a_i(t-s)^2 + (b_i+2 a_i t)(t-s).
 \end{equation}
 A similar representation can be obtained for $y_{i+1}(s) - y_i(s)$
 \begin{align}
 y_{i+1}(s) - y_i(s) &= (a_{i+1} - a_{i}) s^2 + (b_{i+1} - b_i) s + (c_{i+1} - c_i) = A (t-s)^2 + B(t-s) + C, \\
 A &=  a_{i+1} - a_i, \qquad B = 2 A t + b_{i+1} - b_i,  \qquad C = c_i - c_{i+1} + t[b_i - b_{i+1} + (a_i - a_{i+1})t]. \nonumber
 \end{align}
 Same can be done for a polynomial of any degree.

\begin{figure}[!htb]
\begin{center}
\fbox{
\begin{tikzpicture}
\def\lY{0}
\def\rY{12}
\def\stY0{0.3}
\def\stYN{0.8}
\def\angI{10}
\def\angO{5}

\draw[dashed, ultra thick, o-o] (\lY,0,0)--(\rY,0,0)  node[pos=0.5] {$\bullet$};
\draw[dashed, color=green, ultra thick, o-o] (\lY,\stY0,0) to [out=\angI,in=180-\angO]
    node[pos=0.5,draw,fill=green,circle,inner sep=2pt]  (a1) {} (\rY,\stYN,0);
\draw[dashed, color=green, ultra thick, o-o] (\lY,-\stY0,0) to [out=-\angI,in=180+\angO] node[pos=0.5,draw,fill=green,circle,inner sep=2pt] (a2) {} (\rY,-\stYN,0);
\draw[dashed, color=cyan, ultra thick, o-o] (\lY,2*\stY0,0) to [out=2*\angI,in=180-2*\angO] node[pos=0.5,draw,fill=cyan,circle,inner sep=2pt] (a3) {} (\rY,2*\stYN,0);
\draw[dashed, color=cyan, ultra thick, o-o] (\lY,-2*\stY0,0) to [out=-2*\angI,in=180+2*\angO] node[pos=0.5,draw,fill=cyan,circle,inner sep=2pt] (4) {} (\rY,-2*\stYN,0);
\draw[color=blue, ultra thick, o-o] (\lY,4*\stY0,0) to [out=4*\angI,in=180-4*\angO] node[pos=0.5,draw,fill=blue,circle,inner sep=2pt] (a5) {} (\rY,4*\stYN,0);
\draw[color=blue, ultra thick, o-o] (\lY,-4*\stY0,0) to [out=-4*\angI,in=180+4*\angO] node[pos=0.5,draw,fill=blue,circle,inner sep=2pt] (a6) {} (\rY,-4*\stYN,0);

\node at (\rY/2, 3.6){$y_N(t)$};
\node at (\rY/2, -3.6){$y_0(t)$};
\node at (\rY/2, 0.3){$y_i(t)$};
\node at (\rY/2, 1.3){$y_{i+1}(t)$};
\node at (\rY/2, -0.7){$y_{i-1}(t)$};

\node at (\rY/2, 2.8){$\boldsymbol{\vdots}$};
\node at (\rY/2, -2.8){$\boldsymbol{\vdots}$};
\node at (\rY, 2.5){$\boldsymbol{\vdots}$};
\node at (\rY, -2.5){$\boldsymbol{\vdots}$};

\end{tikzpicture}
}
\caption{Internal layers constructed for the given external boundaries $y_0(t)$ and $y_N(t)$, and the number of layers $N$, by using 3 points for each boundary $y_i(t)$ and polynomial curves.}
\label{Bound}
\end{center}
\end{figure}
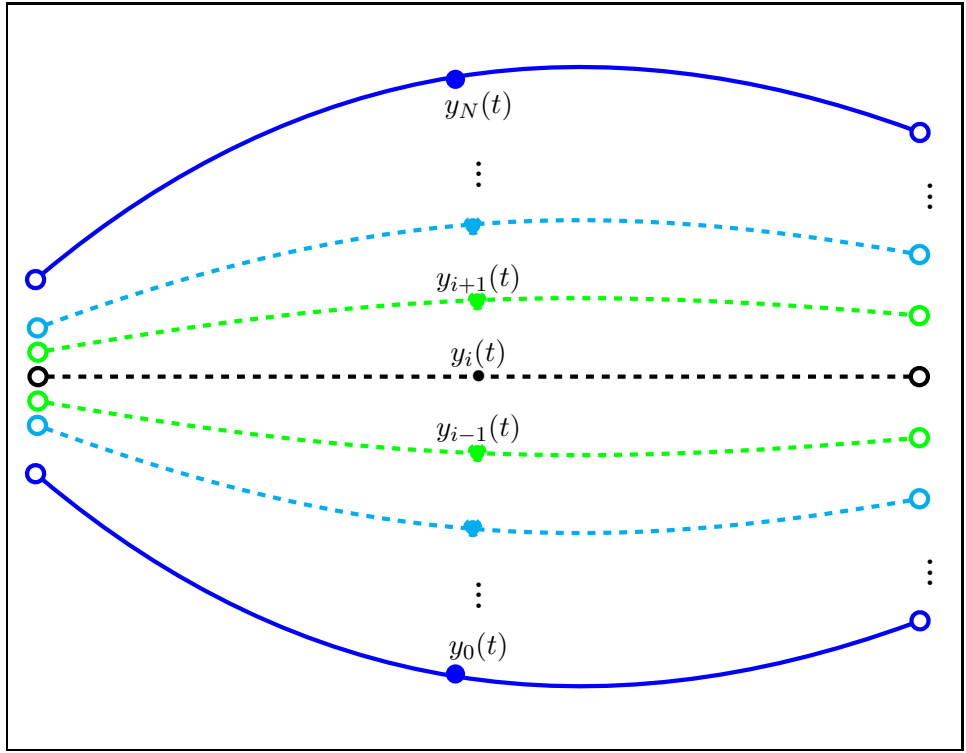

All coefficients $a_i, b_i, c_i$ can be precomputed given the external boundaries. An example of this construction is given in Fig.~\ref{Bound}.

In more detail, suppose that for the given functions $y_1(t) = \chi^-(t), \ y_{N+1} = \chi^+(t)$, we want to have $N$ layers. We use $N$ uniform nodes to split the interval $[\chi^-(0),  \chi^+(0)]$  into $N$ subintervals. We do the same for the interval $[\chi^-(t),  \chi^+(t)]$. If the boundaries $\chi^-(t), \chi^+(t)$ are smooth enough, we can connect points $y_i(0), y_i(t), \ i=2,\ldots,N$ by straight lines in such a way that all boundaries don't cross each other. Suppose this is not possible because the external boundaries are too convex or concave. In that case, we can find some $s = \tau$  where the distance between the external boundaries is minimal and put $N$ nodes there. Then, we can connect all points $y_i(0), y_i(\tau), y_i(t)$ by parabolas, and again check that all the boundaries don't cross each other. We can continue this process by using polynomials of a higher degree to provide the final representation of the boundaries.

Thus, we can find a polynomial of the necessary degree to guarantee that all boundaries don't cross each other unless the external curves have a very peculiar shape, which does not happen in the context of financial applications. Otherwise, we need to use a general approach to the computation of the integrals in \eqref{VolterraTheta}, which will be published elsewhere.

Provided that all the boundaries are constructed in such a way, one can observe that all integrals in \eqref{VolterraTheta} are convolutions. Therefore, we can apply the Laplace transform $\La(f|s,\lambda)$
\begin{equation}
\La(f|s,\lambda) = \int_0^\infty e^{-\lambda s} f(s) ds
\end{equation}
\noindent to both parts of each equation in \eqref{VolterraTheta}. Taking into account that
\begin{align} \label{LT_properties}
 \La(f*g)  &= \La(f) \La(g), \qquad  \La(f') = \lambda\La(f) - f(0),
\end{align}
\noindent and obtain from \eqref{VolterraTheta} a linear system of equations for $\La(\chi_i^+), \La(\chi_i^-), \La(\Omega_i), \La(\Theta_i)$. Using the conditions at the internal boundaries  (such as in \eqref{cont21}), this system can be reduced to a linear system for only $\La(\chi_i^-), \La(\Omega_i)$. One can check that the resulting system is block-diagonal, with all blocks being tridiagonal matrices. Once this system is solved, the functions $\chi_i^-(t), \Omega_i(t)$ is found by using the inverse Laplace transform.

Consider the heat equation in a strip with a piecewise constant thermal conductivity coefficient to illustrate our approach.

\subsection{The heat equation in a strip}

Consider the following problem for the heat equation with a piecewise constant thermal conductivity coefficient (the ML problem):
\begin{align} \label{stat}
\fp{}{x}\left(\sigma^2(x) \fp{U}{x} \right) &= \fp{U}{t}, \qquad (x, t) \in [y_0, y_N] \times \mathbb{R}_+ \\
U(t, y_0) &= 0, \qquad U(t, y_N) = 0. \nonumber \\
U(0, x) &= \delta(x - x_0). \nonumber
\end{align}

Here the thermal conductivity coefficient $\sigma(x)$ is a piecewise constant function of $x$, which changes from layer to layer
\begin{align} \label{sigPWC}
\sigma(x) &= \sum_{i = 1}^{N} \Ind(y_{i-1} < x \le y_{i}) \sigma_i, \\
y_0 &< y_1 < y_2  < ... < y_i <... < y_N. \nonumber
\end{align}
\noindent and $\delta(x)$ denotes the Dirac delta function. As before $y_i$ are the boundaries of the layers in $x$ space. Without loss of generality, we assume that $x_0 \in [y_{j-1}, y_{j}), \ 1 < j < N$.

Due to the initial condition in \eqref{stat} the solution of this problem is Green's function for \eqref{stat}.

We represent the solution $U(t, x)$ in the form
\begin{equation}
U(t, x) = \sum_{i = 1}^{N} \Ind(y_{i-1} < x \le y_{i}) [U_i(t, x) + H_i(t, x)],
\end{equation}
\noindent where the functions $U_i(t, x)$ and $H_i(t, x)$ solve the following problems

\begin{align} \label{UND}
\fp{}{x}\left(\sigma^2_i \fp{U_i}{x} \right) &= \fp{U_i}{t}, \qquad  (x, t) \in (y_{i-1}, y_{i}] \times \mathbb{R}_+, \\
\lim_{x \to y_{i-1}} U_i(t, x) &= \chi_i^-(t), \qquad U_i(t, y_i) = \chi_i^+(t), \nonumber \\
U(0, x) &= 0, \nonumber
\end{align}
\noindent and
\begin{align} \label{HND}
\fp{}{x}\left(\sigma^2_i \fp{H_i}{x} \right) &= \fp{H_i}{t}, \qquad  (x, t) \in (y_{i-1}, y_{i}] \times \mathbb{R}_+, \\
\lim_{x \to y_{i-1}} H_i(t, x) &= 0, \qquad H_i(t, y_i) = 0, \nonumber \\
H(0, x) &= \delta(x - x_0). \nonumber
\end{align}

A well-known physical argument shows that the solution and its flux must be continuous at the layers' boundaries. The first condition yields
\begin{align} \label{cont1}
U_{1} (t, y_0) &= 0 , \\
\lim_{x \to y_i} U_{i} (t, x) &= U_{i+1} (t,y_i), \quad i = 1... N-1, \nonumber \\
\lim_{x \to y_N} U_{N} (t, x) &= 0. \nonumber \
\end{align}

According to \citep{ItkinMuraveyDB}, the function $H(t,x) \ne 0$ only at that interval which contains the point $x_0$, i.e. $[y_{j-1}, y_j)$. Therefore, the flux continuity conditions could be written as
\begin{align} \label{cont2}
\lim_{x \to y_i} \sigma_i^2 \frac{\partial U_i}{\partial x} (t, y_i) &= \sigma_{i+1}^2\frac{\partial U_{i+1}}{\partial x}(t, y_i), \qquad i \neq j-1, j, \\
\lim_{x \to y_{j-1}} \sigma_{j-1}^2 \frac{\partial U_{j-1}}{\partial x} (t, y_{j-1}) &= \sigma_{j}^2\frac{\partial U_{j}}{\partial x}(t, y_{j-1})  + \sigma_j^2 \frac{\partial H_j}{\partial x} (t, y_{j-1}), \nonumber \\
\lim_{x \to y_j} \left[
\sigma_j^2 \frac{\partial U_j}{\partial x} (t, y_j) + \sigma_j^2 \frac{\partial H_j}{\partial x} (t,y_j) \right] &= \sigma_{j+1}^2 \frac{\partial U_{j+1}}{\partial x} (t, y_j). \nonumber
\end{align}

It follows from \eqref{cont1} that
\begin{equation}
\chi_1^-(t) = \chi_N^+(t) = 0, \qquad \chi_i^{+}(t) = \chi_{i+1}^{-}(t), \quad i = 1,\ldots,N-1.
\end{equation}

To simplify the notation let us introduce new functions $f_i(t)$, such as
\begin{equation}
\chi_i^{+}(t) = \chi_{i+1}^{-}(t) = f_i(t), \quad i=0,\ldots,N,
\end{equation}
\noindent so obviously $f_0(\tau) = f_{N+1}(\tau ) = 0$. Using \eqref{VolterraTheta} (or Eq.~3.33 in \citep{ItkinMuraveyDB}), one can get an explicit representation for the derivatives of $U(t,x)$ at each interval
\begin{align} \label{derU}
\dfrac{\partial U_{i}}{\partial x} \Bigg|_{x = y_{i}} &= \frac{f_{i}(t)}{\sigma_{i} \sqrt{\pi t}} - \int_0^t \dfrac{f_{i}(s) - f_{i}(t)}{2\sigma_{i} \sqrt{\pi (t - s)^3}} ds + \int_0^t\left[ f_{i-1}(s) \lambda_{i}^+(t | y_{i-1}, s) - f_{i}(s) \lambda_{0,i}^+ (t | y_{i}, s)\right] ds, \\
\dfrac{\partial U_{i+1}}{\partial x} \Bigg|_{x = y_{i}} &= - \dfrac{f_{i}(t)}{\sigma_{i+1} \sqrt{\pi t}} + \int_0^t
\frac{f_{i}(s) - f_{i}(t)}{2\sigma_{i+1} \sqrt{\pi (t - s)^3}}ds \nonumber
+ \int_0^t\left[ f_{i}(s) \lambda^-_{0,i+1}(t | y_{i}, s) - f_{i+1}(s) \lambda^-_{i+1}(t | y_{i+1}, s)  \right] ds. \nonumber
\end{align}

Here
\begin{align} \label{eta}
\lambda^-_i(t \,|\, \xi ,s) &= \sum_{n = -\infty}^{\infty} \frac{e^{-\frac{(y_i - \xi + 2 n l_i)^2}{4 \sigma_i^2 (t-s)}}} {2\sigma_i^3 \sqrt{\pi (t - s)^3}} \left[ 1 - \frac{(y_i - \xi+ 2 n l_i)^2}{2 \sigma_i^2 (t - s)} \right]
= - \frac{\pi^2}{4 l_i^3} \theta''_3\left[ \frac{\pi(y_i - \xi)}{2 l_i}, q_i(s)\right], \\
\lambda_i^+(\tau \,|\, \xi ,s) &= \sum_{n = -\infty}^{\infty} \frac{e^{-\frac{(y_i - \xi + (2 n+1) l_i)^2}{4 \sigma_i^2 (t -s)}}}{2\sigma_i^3 \sqrt{\pi (t - s)^3}} \left[ 1 - \frac{(y_i - \xi+ (2 n+1) l_i)^2}{2 \sigma_i^2 (t - s)} \right]
= - \frac{\pi^2}{4 l_i^3} \theta''_3\left[ \frac{\pi(y_i + l_i - \xi)}{2 l_i}, q_i(s)\right], \nonumber \\
\lambda^-_{0,i}(t \,|\, \xi, s) &= \lambda^-_{i}(t \,|\,\xi, s)  -  \frac{e^{-\frac{(y_{i-1} - \xi)^2}{4 \sigma_i^2  (t -s)}}}{2\sigma_i^2 \sqrt{\pi (t - s)^3}} \left[ 1 - \frac{y_{i-1} - \xi}{2 \sigma_i^2 (t - s)} \right], \nonumber \\
\lambda^+_{0,i}(t \,|\, \xi, s) &= \lambda^+_{i}(t \,|\,\xi, s) - \frac{e^{-\frac{(y_i - \xi + l_i)^2}{4\sigma_i^2 (t -s)}}}{2\sigma_i^2\sqrt{\pi (t - s)^3}} \left[ 1 - \frac{(y_i - \xi+ l_i)^2}{2\sigma_i^2(t - s)} \right]. \nonumber \\
q_{i}(s) &= e^{- \pi^2 \sigma_{i}^2 (t-s)/l_{i}^2}. \nonumber
\end{align}
Here $\theta_i(z,p), \ i=2,3$ are the Jacobi theta functions of the second and third kind, \citep{mumford1983tata}, and $\theta''_3(z,p)$ is the second derivative of $\theta_3(z,p)$ on the first argument.

Substituting $\xi = y_{i-1}, y_i, y_{i+1}$ into \eqref{eta} we obtain
\begin{align} \label{tr1}
\lambda^-_{i+1}(t \,|\, y_{i+1},s) &= - \frac{\pi^2}{4 l_{i+1}^3} \theta''_3(0, q_{i+1}(s)), \\
\lambda_i^+(\tau \,|\, y_{i-1} ,s) &= - \frac{\pi^2}{4 l_i^3} \theta''_3(\pi, q_i(s)), \nonumber \\
\lambda^-_{0,i+1}(t \,|\, y_{i}, s) &=\sum_{\substack{n = -\infty\\ n \neq 0}}^{\infty}
\frac{ e^{-\frac{(2 n l_{i+1})^2} {4 \sigma_{i+1}^2 (t-s)}}} {2\sigma_{i+1}^3 \sqrt{\pi (t - s)^3}}
\left[ 1 - \frac{(2 n l_{i+1})^2}{2 \sigma_{i+1}^2 (t - s)} \right], \nonumber \\
\lambda^+_{0,i}(t \,|\, y_{i}, s) &= - \frac{\pi^2}{4 l_i^3} \theta''_3(\pi, q_i(s))
- \frac{e^{-\frac{l_{i}^2}{4\sigma_{i}^2 (t -s)}}}{2\sigma_{i}^2\sqrt{\pi (t - s)^3}} \left[ 1 - \frac{l_{i}^2}{2\sigma_{i}^2(t - s)} \right]. \nonumber
\end{align}

 Since at $s \to t$ we have $q(s) \to 1$, all RHSs in \eqref{tr1} are regular in this limit and vanish.  The latter is due to the fact that $\lim_{q \to 1} \theta''_3 (0,q) = \theta''_3 (\pi, q) = 0$.  Therefore, all integral kernels in \eqref{derU} are regular. We also assume that functions $f_i(s)$ are smooth enough, so that in the limit all the integrals vanish.

Applying integration by parts to \eqref{derU}, we get the following simplified system
\begin{align}
\frac{\partial U_i}{\partial x} \Bigg|_{x = y_{i-1}} = &-\int_0^t\left[ \eta^{even}_i (t, s) d\left(f_{i-1}(s)\right)
- \eta^{odd}_i (t, s) d\left(f_{i}(s)\right)  \right] \\
&-\frac{f_{i-1}(0)}{\sigma_i \sqrt{\pi t}} - f_{i-1}(0) \eta_i^{even} (t,0) + f_{i}(0) \eta_i^{odd}(t, 0),
 \nonumber \\
\frac{\partial U_i}{\partial x} \Bigg|_{x = y_i} = &-\int_0^t\left[ \eta^{odd}_i (t, s) d\left(f_{i-1}(s)\right)
- \eta^{even}_i (t, s) d\left(f_{i}(s)\right)  \right] \nonumber \\
&+\frac{f_{i}(0)}{\sigma_i \sqrt{\pi t}} + f_{i}(0) \eta_i^{even} (t,0) - f_{i-1}(0) \eta_i^{odd}(t, 0), \nonumber
\end{align}
\noindent where
\begin{eqnarray} \label{etaDef}
\eta_i^{even}(t-s) =  \frac{1}{\sigma_i \sqrt{\pi (t - s)}}\sum_{n= -\infty}^{\infty} e^{-\frac{(2n l_i)^2}{4\sigma_i^2(t -s)}}, \quad
\eta_i^{odd}(t-s) = \frac{1}{\sigma_i \sqrt{\pi (t - s)}} \sum_{n= -\infty}^{\infty} e^{-\frac{((2n+1) l_i)^2}{4\sigma_i^2(t -s)}}.
\end{eqnarray}

Indeed, since $\eta^-_i(t | y_{i - 1}, t) = 0, \ \eta^-_i(t | y_{i}, t) = 0, \ f_{i-1}(0) = 0, \ f_{i}(0) = 0$ we have
\begin{align}
&- \frac{f_{i-1}(t)}{\sigma_i \sqrt{\pi t}} + \int_0^t \frac{f_{i-1}(s) - f_{i-1}(t)}{2\sigma_i \sqrt{\pi (t - s)^3}} + \int_0^t\left[ f_{i-1}(s) d\left(\eta^-_i(t | y_{i-1}, s)\right) -f_{i}(s) d\left(\eta^-_i(t | y_{i}, s)\right)  \right] \\
&= -\frac{f_{i-1}(t)}{\sigma_i \sqrt{\pi t}} -\frac{f_{i-1}(t)}{\sigma_i \sqrt{\pi (t - s)}} \Bigg|_{s = 0}^{s = t} + \frac{f_{i-1}(s)}{\sigma_i \sqrt{\pi (t - s)}} \Bigg|_{s = 0}^{s = t} - \int_0^t \frac{1}{\sigma_i \sqrt{\pi (t -s)}} d\left(f_{i-1} (s) \right) + \nonumber \\
&+ f_{i-1}(s) \eta^-_i(t | y_{i-1}, s)\Bigg|_{s = 0}^{s = t} - f_{i}(s) \eta^-_i(t | y_{i}, s)\Bigg|_{s = 0}^{s = t} -\int_0^t\left[\eta^-_i(t | y_{i-1}, s) d\left( f_{i-1}(s)\right) - \eta^-_i(t | y_{i}, s)d\left(f_{i}(s)\right)  \right]
\nonumber \\
&= -\int_0^t\left[\left(\eta^-_i(t | y_{i-1}, s) + \frac{1}{\sigma_i \sqrt{\pi (t -s)}} \right) d\left( f_{i-1}(s)\right)
- \eta^-_i(t | y_{i}, s)d\left(f_{i}(s)\right)  \right] \nonumber \\
&= -\int_0^t\left[ \eta^{even}_i (t, s) d\left(f_{i-1}(s)\right) - \eta^{odd}_i (t, s) d\left(f_{i}(s)\right)  \right]. \nonumber
\end{align}

In turn, as shown in \citep{ItkinMuraveyDB}, Eq.~3.31, the function $H_j(t, x)$ can be represented  as follows
\begin{equation}
H_j(t, x) = \frac{1}{2 \sigma_j \sqrt{\pi t}} \sum_{n = -\infty} ^{\infty} \left[e^{- \frac{(2n l_j +  x- x_0)^2}{4 \sigma_j^2 t  }} - e^{- \frac{(2n l_j +  x+ x_0 - 2y_{j-1})^2}{4 \sigma_j^2 t}} \right].
\end{equation}

Hence, the gradients at the boundaries are
\begin{align} \label{tr2}
\frac{\partial H_j(t, x)}{\partial x}\Bigg|_{x = y_{j-1}} &\equiv \upsilon^-(t | x_0, 0) \\
&= \fp{}{x} \left\{ \frac{1}{2 l_j}\theta_3 \left[\frac{\pi(x - x_0)}{2 l_j}, q_j(0) \right] -
\frac{1}{2 l_j}\theta_3 \left[\frac{\pi(x + x_0 - 2 y_{j-1})}{2 l_j}, q_j(0) \right] \right\} \Bigg|_{x=y_{j-1}} \nonumber \\
&= \frac{1}{4 l_j^2} \left\{\theta'_3 \left[\frac{\pi(y_{j-1} - x_0)}{2 l_j}, q_j(0) \right]
- \theta'_3 \left[\frac{\pi(x_0 - y_{j-1})}{2 l_j}, q_j(0) \right] \right\}, \nonumber \\
\frac{\partial H_j(t, x)}{\partial x}\Bigg|_{x = y_{j}} &= \frac{1}{4 l_j^2} \left\{\theta'_3 \left[\frac{\pi(y_{j} - x_0)}{2 l_j}, q_j(0) \right] - \theta'_3 \left[\frac{\pi(x_0 - y_{j})}{2 l_j}, q_j(0) \right] \right\}. \nonumber
\end{align}

Thus, we obtain the following system of Volterra equations
\begin{align} \label{Volterra_system}
\int_0^t \Bigg[ &-\sigma_i^2 \eta^{odd}_i (t-s) d\left(f_{i-1}(s)\right) +  \left(\sigma_i^2\eta^{even}_i (t-s) + \sigma_{i+1}^2\eta^{even}_{i+1} (t- s)  \right) d\left(f_{i}(s)\right) \\
&-\sigma_{i+1}^2 \eta^{odd}_{i+1} (t-s) d\left(f_{i+1}(s)\right) \Bigg] =h_{i}(t), \nonumber \\
h_{i}(t) &= 0, i \neq j-1, j, \qquad h_{j-1}(t) = \sigma_j^2\upsilon^-(t | x_0, 0), \qquad
h_{j}(t) = -\sigma_j^2\upsilon^+(t | x_0, 0). \nonumber
\end{align}

Since the kernels depend only on $t -s$ one can rewrite the above equations as a convolution
\begin{equation} \label{VolTrans}
\left(-\sigma_i^2 \eta_i^{odd}(\cdot) * f'_{i-1} (\cdot)  + \left[\sigma_i^2 \eta_i^{even}(\cdot) +\sigma_{i+1}^2 \eta_{i+1}^{even}(\cdot) \right]* f'_{i} (\cdot)  - \sigma_{i+1}^2 \eta_{i+1}^{odd}(\cdot) * f'_{i+1} (\cdot) \right) (t) = h_i(t).
\end{equation}

\subsection{The Laplace transform}

Applying the Laplace transform to \eqref{VolTrans}, we get
\begin{equation}
\sqrt{\lambda} \left(-\sigma_i^2  \La(\eta_i^{odd}) \, \La(f_{i-1}) +
\left(\sigma_i^2 \La(\eta_i^{even}) + \sigma_{i+1}^2 \La(\eta_{i+1}^{even}) \right)\La(f_{i}) - \sigma_{i+1}^2  \La(\eta_{i + 1}^{odd}) \, \La(f_{i+1}) \right) = \frac{\La(h_i)}{\sqrt{\lambda}},
\end{equation}
\noindent or, in the matrix form
\begin{equation} \label{Msystem}
\textbf{M} \textbf{g} = \frac{\sigma_j^2}{\sqrt{\lambda}} \left[ \La(\upsilon^-(t|x_0, 0)) \Ind_{j-1}  - \La( \upsilon^+(t|x_0, 0)) \Ind_{j} \right].
\end{equation}	

Here $\Ind_{j}$ denotes the indicator vector, i.e.
\begin{equation} 	\label{IndVecDef}
\Ind_{j} = \left(\underbrace{0, 0, ..0}_{j-1},1, 0, ... 0\right)^\top,
\end{equation}
\noindent the vector $\textbf{g}$ is the column vector
\begin{equation} \label{gvec_def}
\textbf{g} = \left(\La f_1, \ldots,\La f_{N-1} \right)^\top,
\end{equation}
\noindent and the matrix $\mathbf{M}$ is a symmetric tridiagonal matrix
\begin{eqnarray} \label{matM}
\mathbf{M} =
\begin{pmatrix}
D_1 & -\beta_1&   &  &   \\
 -\beta_1& D_2 & -\beta_2  &  & \\
 &  -\beta_2& \ddots & \ddots & \\
 &  &  \ddots&  \ddots  & -\beta_{N-2} \\
  &  &   & -\beta_{N-2}&  D_{N-1}
\end{pmatrix},
\end{eqnarray}

Coefficients of the matrix $\mathbf{M}$ have the form
\begin{equation} \label{betaD_def}
\beta_i = \sqrt{\lambda}\sigma^2_{i+1} \La(\eta_{i+1}^{odd}), \qquad D_i =  \sqrt{\lambda}\left[ \sigma^2_i \La( \eta_i^{even}) + \sigma^2_{i+1} \La(\eta_{i+1}^{even})\right],
\end{equation}
\noindent and can be found explicitly, see Appendix~\ref{App3}
\begin{alignat}{2}
\La(\eta_{i}^{even}) &= \frac{1}{\sigma_i \sqrt{\lambda}} \coth\left( \frac{\sqrt{\lambda} l_i}{\sigma_i}\right), \qquad & \La(\eta_{i}^{odd}) &= \frac{1}{\sigma_i \sqrt{\lambda}} \frac{1}{\sinh\left( \frac{\sqrt{\lambda}  l_i} {\sigma_i}\right)}, \\
\La(\upsilon^-(t |x_0, 0)) &=  \frac{1}{\sigma_j^2}  \frac{\sinh\left(\frac{(y_{j+1} - x_0) \sqrt{\lambda}} {\sigma_j}\right)} {\sinh\left( \frac{l_j \sqrt{\lambda}}{\sigma_j} \right)},
\qquad  &\La(\upsilon^+(t |x_0, 0)) &= -\frac{1}{\sigma_j^2} \frac{\sinh\left(\frac{(x_0-  y_{j})  \sqrt{\lambda}} {\sigma_j}\right)} {\sinh\left( \frac{l_j \sqrt{\lambda}}{\sigma_j} \right)}. \nonumber
\end{alignat}

Finally, introducing the notation
\begin{equation} \label{omega_gamma_def}
\omega_i = \frac{l_i}{\sigma_i}, \qquad \gamma_1 = \frac{y_{j+1} - x_0}{l_j}, \quad \gamma_2 = \frac{x_0- y_j}{l_j},
\end{equation}
\noindent the system \eqref{Msystem} can be represented as
\begin{align} \label{Msystem_final}
\mathbf{M}\mathbf{g} &= \frac{1}{\sqrt{\lambda}} \left[
\frac{ \sinh\left(\gamma_1 \omega_j \sqrt{\lambda}\right) } { \sinh\left( \omega_j \sqrt{\lambda}\right) }  \Ind_{j-1} + \frac{ \sinh\left(\gamma_2 \omega_j \sqrt{\lambda}\right) } { \sinh\left( \omega_j \sqrt{\lambda}\right) } \Ind_{j} \right], \\
D_{i} &= \sigma_i \coth \left(\omega_i \sqrt{\lambda}\right) + \sigma_{i+1} \coth \left(\omega_{i+1} \sqrt{\lambda}\right),
\qquad \beta_i = \frac{\sigma_{i+1} }{\sinh\left(\omega_{i +1}\sqrt{\lambda} \right)}. \nonumber
\end{align}

\section{Numerical experiments} \label{results}

In this section, we solve the problem in \eqref{stat} by using the ML method. Note that such problems appear both in physics and in finance. A simple example of a financial problem is finding Green's function for pricing double barrier options written on the underlying $S_t$ with local volatility  \begin{equation}
d S_t = \sigma(S_t) d W_t,
\end{equation}
\noindent where $\sigma(S)$ is a piecewise constant function. Below we describe two numerical experiments.

\subsection{Constant volatility $\sigma_i$}

To start with, we assume that $\sigma_i = const, \ i=1,\ldots,N$, and hence, $\sigma_i^2$ in \eqref{stat} can be pulled out of the derivative in $x$. This problem has an analytic solution, see \citep{Lipton2001} and references therein, which can be represented as the Fourier series. Re-writing it by using the definition of Jacobi theta functions yields
\begin{align}
U(T,y) &= \frac{1}{2 l} \left[ \theta_3\left( \frac{\pi (y-x_0)}{2 l}, q\right) -
\theta_3\left( \frac{\pi (y + x_0 - 2 y_0)}{2 l}, q\right)\right], \\
l &= y_N - y_0, \quad q = e^{- \frac{\pi^2 \sigma^2}{l^2}T}. \nonumber
\end{align}

To solve this problem, we need first to solve the linear system in \eqref{Msystem_final} numerically, and then use the corresponding Laplace images to find the function $f_i = f(y_i), \ i=1,\ldots,N$ by applying an inverse Laplace transform. For the latter step, we use the Gaver-Stehfest method
\begin{equation} \label{GSA}
f(T,y) = \Lambda \sum_{s=1}^{[m]} St_s^{(m)} \La(f_i(\Lambda)), \quad \Lambda = \dfrac{\log 2}{T}.
\end{equation}
This algorithm was widely studied (see, e.g., \citep{Kuznetsov2013} and references therein), and, provided that the resulting function is non-oscillatory, converges very quickly. For instance, choosing $m=12$ terms in the series representing the solution is usually sufficient. The coefficients $St_s$ can be found explicitly in advance.
\begin{table}[!htb]
\begin{center}
\caption{Parameters of the test.}
\label{tab1}
\begin{tabular}{|c|c|c|c|c|c|c|}
\hline
$y_0$ & $y_N$ & $\sigma$ & $T$ & $N$ & $m$ \\
\hline
-1.0 & 1.0 & 0.5 & 1.0 & 20  & 16\\
\hline
\end{tabular}
\end{center}
\end{table}

\begin{figure}[H]
\begin{center}
\subfloat[]{%
\label{Exp1-a}
\fbox{\includegraphics[width=0.48\textwidth]{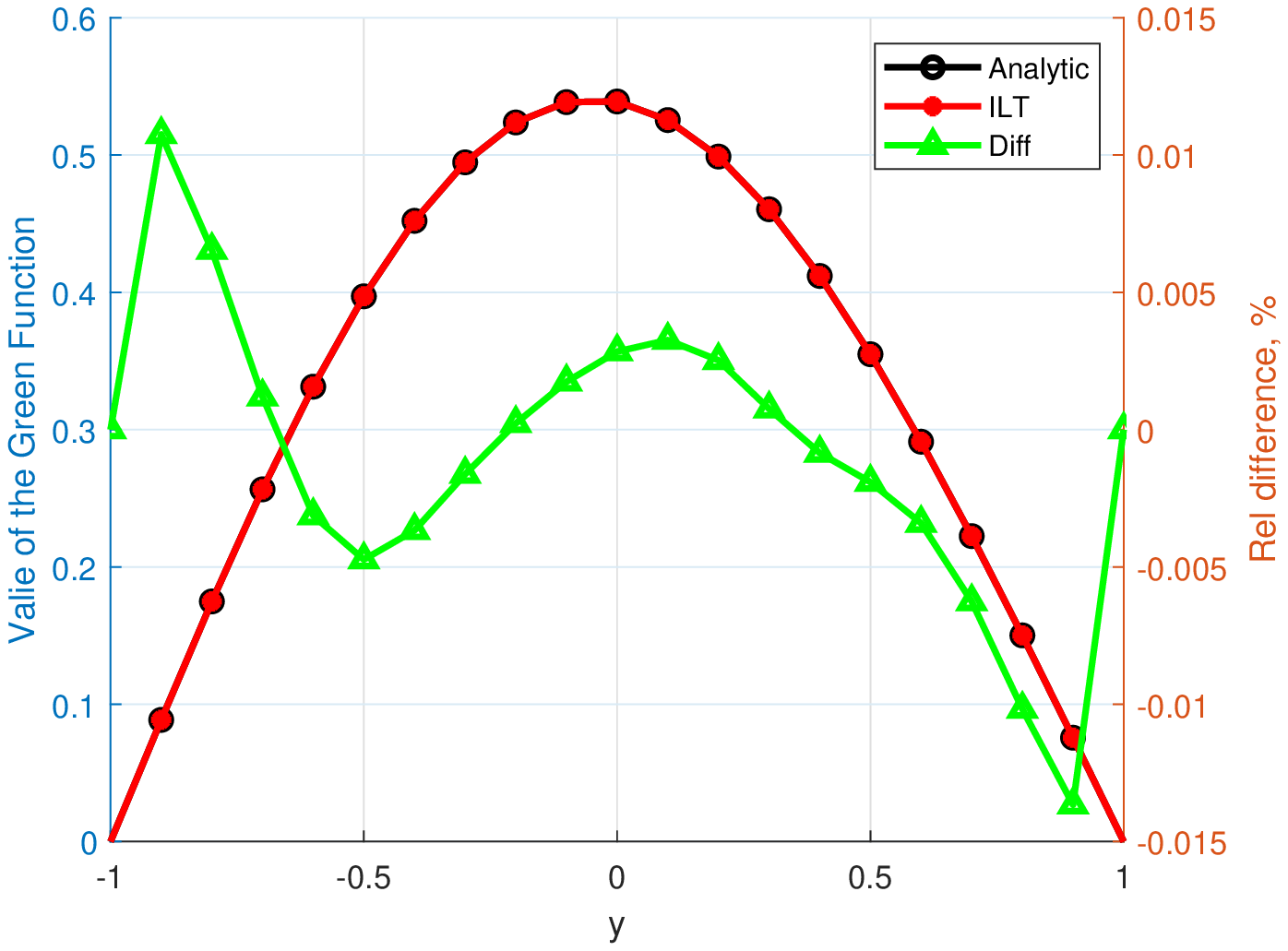}}
}
\subfloat[]{%
  \fbox{\includegraphics[width=0.48\textwidth]{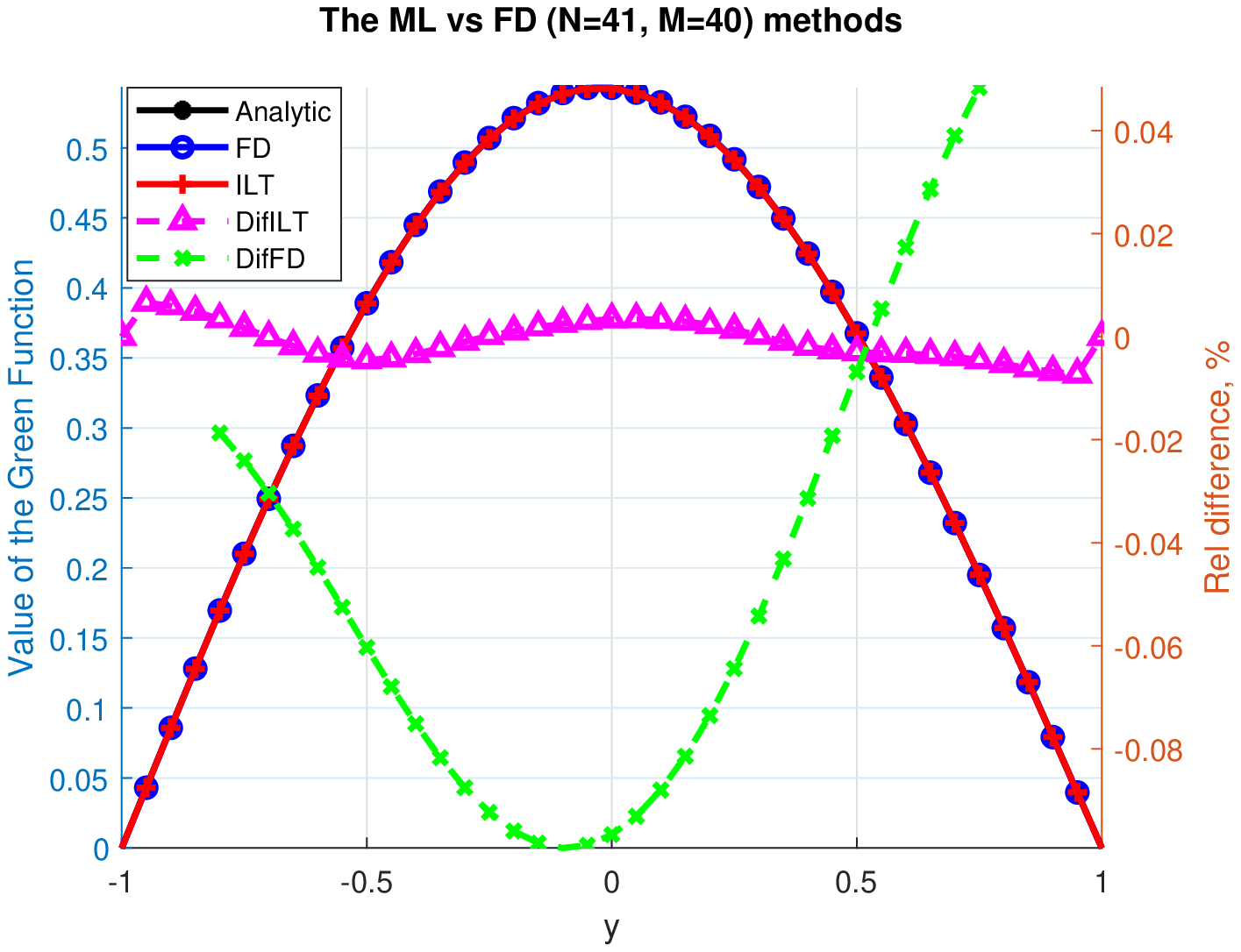}}
\label{Exp1-b}
}
\caption{Comparison of the Analytic and ML solutions (a), and Analytic, ML and FD solutions (grid with $41\times 40$ nodes) (b) for $\sigma_i = 0.5, T = 1$. Here {\it Analytic} denotes the analytic solution of the problem, {\it ILT} - the ML solution, {\it FD} - the FD solution, {\it DiffILT} - the relative error of the ML solution with respect to the analytic one, {\it DifFD} - same for the FD method.}
\label{Exp1Fig}
\end{center}
\end{figure}

The model parameters for this test are given in Table~\ref{tab1}, and the results are depicted in Fig.~\ref{Exp1-a}. Here, the left vertical axis shows the values of $y_i(T)$, and the right vertical axis shows the relative error (in percent) of the solution compared with the analytic one.

The ML solution coincides with the analytic one with high accuracy. The elapsed time of the experiment is 8.8 ms (we run our code, written in Matlab, on a PC with two Quadcore CPU Intel i7-4790 3.60 GHz). The elapsed time doesn't depend on the option maturity $T$, so the calculation is fast even for long maturities. Note that the ML solution's computation takes only 2.3 ms, while the remaining time is used for computing the Gaver-Stehfest coefficients (but those can be precomputed if so desired).
Since the matrix $\mathbf{M}$ is tridiagonal, the ML method's complexity is $O(m N)$. By comparison, the complexity of the finite difference (FD) method is $O(M N)$, $M$ is the number of time steps. Obviously, for long maturities $M \gg m$, so the FD method is slower.

To validate this, we also implemented an FD method to solve the same problem. The FD solver runs on a uniform grid and is a Crank-Nicolson scheme after four steps, while for the first four steps it uses an implicit Euler scheme. In other words, we start with four Ranacher steps, see \citep{ItkinBook} and references therein.

To have the same spatial approximation in $y$, we need to run both the ML and FD methods with the same number of nodes. While the ML method provides an accurate result even at $N=10$, the FD method fails and needs at least 40 nodes to converge to the solution. Therefore, we choose $N=41$. The same is true for the time step, so the minimal number of FD time steps in our experiment is $M=40$, while the ML method provides the solution at $t=T$ just at once. The results of the comparison of both methods with the analytic solution are given in Fig.~\ref{Exp1-b}. Again, the left vertical axis shows the values of $y_i(T)$, and the right vertical axis is the relative error (in percent) of the solution compared with the analytic one.

It is clear that the accuracy of the FD method is worse than that of the ML method. By increasing $N$ and $M$, one can improve the FD method's accuracy, but it takes time. The elapsed time for the FD method with $N=41, M=40$ is 41 ms, so it is about 18 times slower than the ML method. Obviously, for longer maturities, more time-steps are necessary, so the FD method becomes even slower.

\begin{figure}[!htb]
\begin{center}
\subfloat[]{%
\label{Exp1s-a}
\fbox{\includegraphics[width=0.48\textwidth]{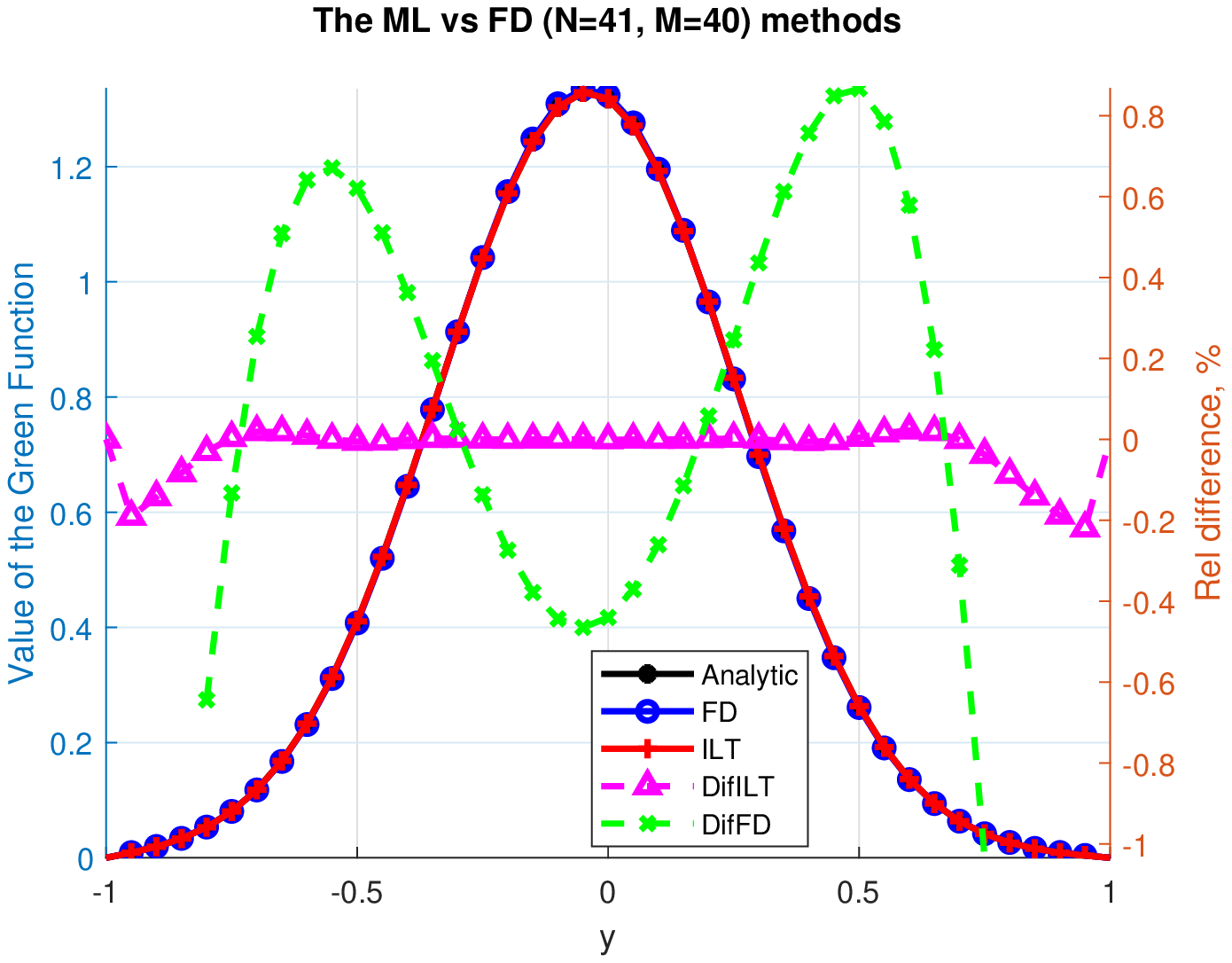}}
}
\subfloat[]{%
  \fbox{\includegraphics[width=0.48\textwidth]{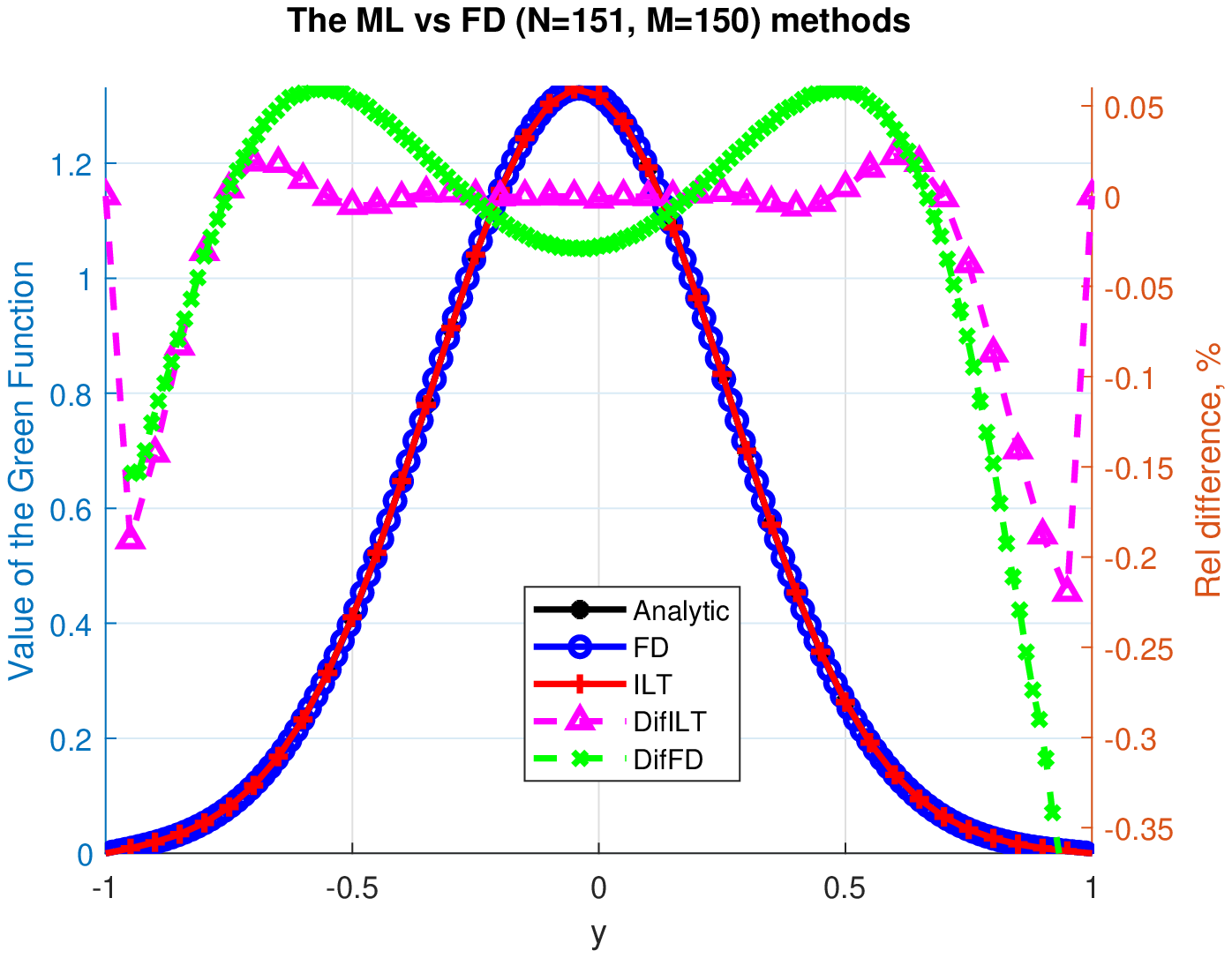}}
\label{Exp1s-b}
}
\caption{Comparison of the Analytic, ML and FD solutions for $\sigma_i = 0.3, T = 0.5$. Here {\it Analytic} denotes the analytic solution of the problem, {\it ILT} - the ML solution, {\it FD} - the FD solution, {\it DiffILT} - the relative error of the ML solution with respect to the analytic one, {\it DifFD} - same for the FD method.}
\label{Exp1small}
\end{center}
\end{figure}

It is also known that for small $T$ and volatilities, the FD method's error increases. To illustrate this fact, we run the same experiment, but now with $T=0.5, \ \sigma = 0.3$. The results are presented in Fig.~\ref{Exp1small}.

Fig.~\ref{Exp1s-a} shows that the accuracy of the ML method is still good with the same number of internal layers $N=40$, while the error of the FD method with $N=41, M=40$ is quite significant. The error can be reduced by running the FD method with $N=151, M=150$ (see Fig.~\ref{Exp1s-b}); however, the corresponding elapsed time is 86 ms. Thus, for the same accuracy, the ML method is 37 times faster.

\subsection{Piecewise constant volatility $\sigma_i$}

Here $\sigma(x)$ is a piecewise constant function defined in \eqref{sigPWC}. In this case, there is no analytic solution of the problem\footnote{Based on \eqref{Msystem_final} it is possible to derive an explicit series representation of the solution. It will be published elsewhere.}, hence as a benchmark, we use an FD method, namely the same FD solver as in the previous experiment.  However, since we solve the problem in \eqref{stat}, an FD scheme has to be implemented for the conservative heat equation. While such an implementation is possible, we prefer to rewrite \eqref{UND} and \eqref{HND} in a non-divergent form. For instance,
\begin{align}
\fp{U_i}{t} &= \Xi_i^2(x) \sop{U_i}{x}  + \fp{\Xi^2(x) }{x} \fp{U_i}{x}, \\
\Xi_i(x) &= \sigma_i [\Theta_H (x_{i+1}-x) - \Theta_H (x_{i}-x)], \nonumber
\end{align}
\noindent where $\Theta_H(x)$ is the Heaviside theta function, \citep{as64} with $\Theta_H(0) = 1$. Accordingly, on the interval $x \in (y_i, y_{i+1}]$ we have
\begin{equation}
\fp{\Xi^2(x) }{x} = \sigma^2_{i+1}[\delta(x_i - x) - \delta(x_{i+1} - x)].
\end{equation}
At the point $x=x_{i+1}$ this gives $\partial \Xi^2(x)/\partial x = - \sigma^2_{i+1} \delta(0)$. In turn, $\delta(0)$ can be numerically approximated as
\begin{equation} \label{d0}
\delta(0) = \frac{2}{y_N - y_0},
\end{equation}
\noindent which provides the correct normalization of the Dirac delta function. Indeed, the integral over the interval $[y_0, y_n]$ of the test  function equal to $1$ at $x=y_{i+1}$ and $0$ otherwise  computed by using a trapezoidal rule is equal to $(y_N - y_0)/2$. Therefore, we need to use \eqref{d0} to provide the correct numerical normalization.
\begin{table}[!htb]
\begin{center}
\caption{Parameters of the second experiment.}
\label{tab2}
\begin{tabular}{|c|c|c|c|c|c|c|}
\hline
$y_0$ & $y_N$ & $T$ & $N$ & $m$ & M \\
\hline
-1.0 & 4.0 & 2.0 & 50  & 16 & 100\\
\hline
\end{tabular}
\end{center}
\end{table}

In this experiment, we use parameters of the model given in Table~\ref{tab2}, and the piecewise constant volatility $\sigma_i$, which is defined as follows
\begin{equation} \label{sigmaPWC}
\sigma_i(s) = e^{-i/N}, \quad s \in (y_i, y_{i+1}], \ i=1,\ldots,N.
\end{equation}

\begin{figure}[H]
\begin{center}
\subfloat[]{%
\label{Exp2-a}
\fbox{\includegraphics[width=0.48\textwidth]{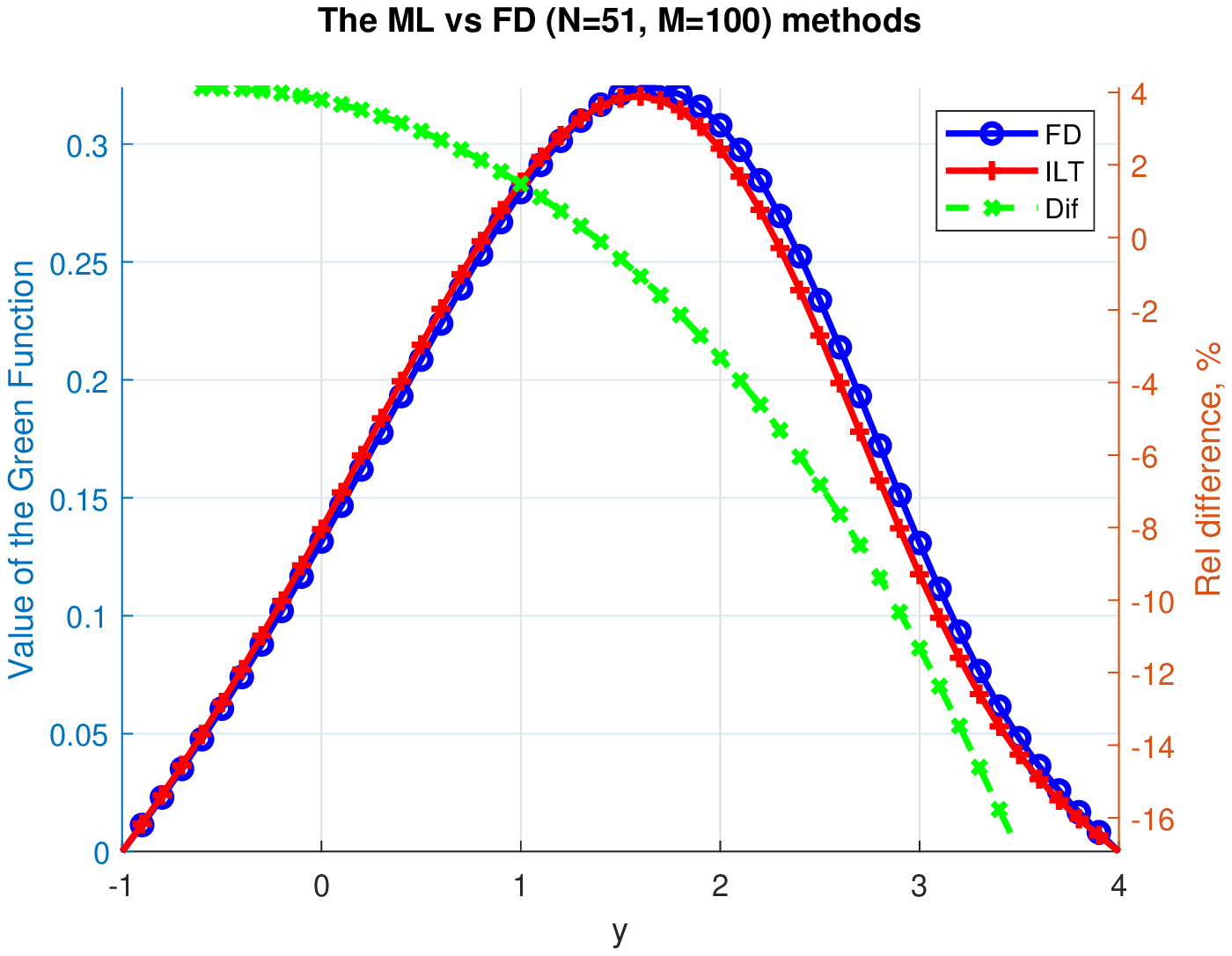}}
}
\subfloat[]{%
  \fbox{\includegraphics[width=0.48\textwidth]{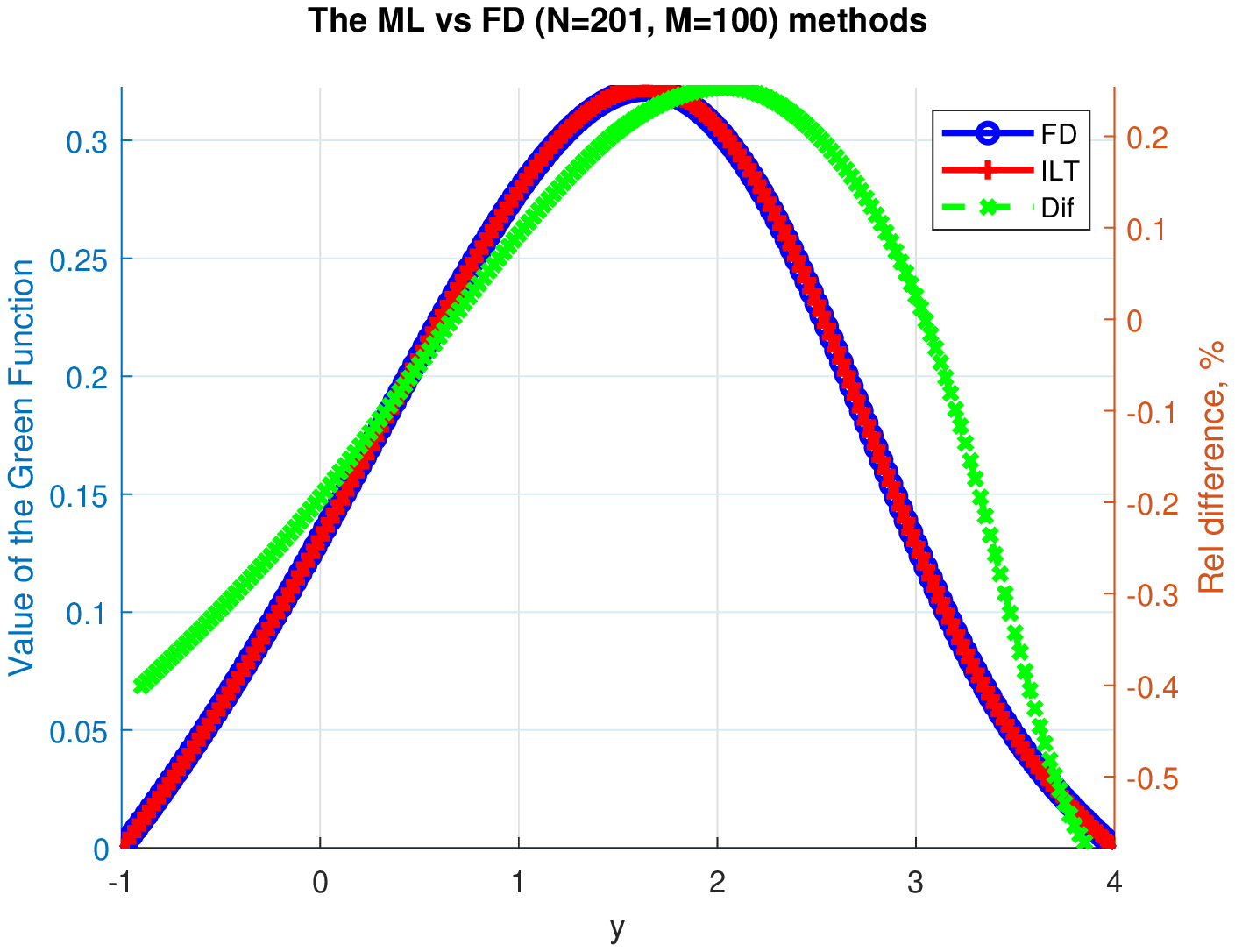}}
\label{Exp2-b}
}
\caption{Comparison of the  ML and FD solutions for a piecewise constant $\sigma(x)$. Here {\it ILT} denotes the ML solution, {\it FD} - the FD solution, {\it Dif} - the relative error of the FD solution with respect to the ML one.}
\label{Exp1pwc}
\end{center}
\end{figure}

The results of the test are presented in Fig.~\ref{Exp2-a}. Again, the number of nodes for the ML and FD methods is the same. The difference between the two solutions reaches 16\% at the right external boundary, 4\% at the left external boundary, and changes in this range in between. The elapsed time is 2.2 ms for the ML and 50 ms for the FD methods.

To check the convergence of the solution, we rerun the calculation with $N=200$. The results are presented in Fig.~\ref{Exp2-b}.
The relative error $\varepsilon$ drops down to be in $\varepsilon \in [-0.6, 0.2]$ percent. The elapsed time is 5 ms for the ML and 88 ms for the FD methods. Note that in this case, to reduce the error, we also need to increase the number of layers for the ML method. This effect is explained in Section~\ref{Discus}; it is due to the non-smoothness of $\sigma$ in this experiment.

The physics meaning of the obtained results is as follows. Suppose we consider diffusion rather than heat conduction. According to \eqref{sigmaPWC} the diffusion coefficient $\sigma^2(x)$ is a decreasing function of $x$ when moving from $y_0$ to $y_N$. Since we request continuity of the flux at the internal boundaries, the gradient of the solution increases with $x$ when moving from left to right. The maximum of the solution, which is located at $x=x_0$ when $t=0$, travels to the right when $t$ increases. Recall that the solution is the Green function of our problem. This behavior was also observed in \citep{Lan_on_2001}, where the authors studied particles trapped between two nearly parallel walls making their confinement position dependent. They not only measured a diffusion coefficient which depended on the particles’ position but also reported and explained a new effect: a drift of the particles individual positions (so change in concentration) in the direction of the diffusion coefficient gradient, in the absence of any external force or concentration gradient.

\section{Discussion} \label{Discus}

In the previous section, we have demonstrated that the ML method's complexity is linear in $N$. The same is true in the general case because, as mentioned at the end of Section ~\ref{solVolt}, from  \eqref{VolterraTheta} we obtain a linear system of equations for $\La(\chi_i^-), \La(\Omega_i)$ which has a block diagonal matrix with all blocks being tridiagonal matrices. Therefore, the ML method's complexity remains linear and approximately is $O(4 m N)$, so it doesn't depend on $T$. Hence, if $4 m$ is of the order of $M$, the ML and the FD methods have the same complexity. For typical values $m=12$ we have $4m = 48$. Therefore, for short maturities $T < 1$ year, both methods' complexity is roughly the same. However, our method has an obvious advantage for the long maturities occurring in the Fixed Income context.

The ML method has some other advantages as well. First, the FD construction provides only the values of the unknown function at the grid nodes in space, and at intermediate points they can be found only by interpolation. In contrast, using the ML method, we obtain an analytic representation of the solution at any $x$ (once the values at the layers' boundaries are found). Second, the Greeks, i.e., derivatives of the solution, can be expressed semi-analytically by differentiating the solution with respect to $x$ or some  parameter of the model and performing numerical integration, provided that the values at the internal boundaries are found. For the FD method, the Greeks can be found only numerically. Moreover, to compute the Vega, a new run of the FD method is required, while for the ML method, all Greeks can be calculated in one go, as described.

As far as an approximation with respect to $x$ is concerned, the following observation holds. Using the ML method, we obtain an analytical solution at every interval $i, \ i=1,\ldots,N$. However, to do this, we need to approximate the corresponding coefficient, e.g., $\sigma(x)$ over layers by piecewise constant or linear functions. For the linear approximation, the solution's accuracy is $O((\Delta x)^2)$, i.e., same as for the FD method of the second order. Therefore, it seems that the spatial accuracies of both ML and FD methods are the same.

On the other hand, the error of both methods is also proportional to the second derivative. For the FD method, this is the second derivative of the solution; for the ML method - the second derivative of the coefficient, e.g., $\sigma_{x,x}(x)$. If the latter is smaller than the second derivative of the solution (say, the option Gamma), then the number of layers $N$ can be decreased while providing the same accuracy. This reduction provides an additional speedup of our method as compared with the FD method. This fact is illustrated by our first experiment where function $\sigma(x)$ is smooth, so even a small number of layers is sufficient to obtain a very accurate solution. In the second experiment, $\sigma(x)$ jumps at the layer's boundaries, and, therefore, one needs to increase the number of layers to provide the same accuracy.

Note that for the FD method, the difficulties caused by sharp gradients can be alleviated by using nonuniform grids where the nodes are condensed in the area where gradients are high. The same approach could be applied to the construction of internal layers in the ML method.

Overall, we can conclude that the new ML method proposed in this paper is significantly faster than the FD method, provides better accuracy, and represents the solution in a semi-analytical form. The method's speed is close to that for the Radial Basis Functions (RBF) approach, \citep{YCHon3,Fasshauer2,Pettersson}, while other properties listed above are superior to the RBF.

\section*{Acknowledgments}

We are grateful to Peter Carr for some fruitful discussions. Dmitry Muravey acknowledges support by the Russian Science Foundation under the Grant number 20-68-47030.

%%%%%%%%%%%%%%%%%%%%%%%%%%%%%%%%%%%%%%%%%%%%%%%%%%%%%%%%%%%%
%\printbibliography[title={References}]
%\bibliographystyle{plainnat}
%\bibliography{tdOU,localVG}

\begin{thebibliography}{47}
\providecommand{\natexlab}[1]{#1}
\providecommand{\url}[1]{\texttt{#1}}
\expandafter\ifx\csname urlstyle\endcsname\relax
  \providecommand{\doi}[1]{doi: #1}\else
  \providecommand{\doi}{doi: \begingroup \urlstyle{rm}\Url}\fi

\bibitem[Abramowitz and Stegun(1964)]{as64}
M.~Abramowitz and I.~Stegun.
\newblock \emph{Handbook of Mathematical Functions}.
\newblock Dover Publications, Inc., 1964.

\bibitem[Andersen and Piterbarg(2010)]{andersen2010interest}
L.B.G. Andersen and V.V. Piterbarg.
\newblock \emph{Interest Rate Modeling}.
\newblock Number v. 2 in Interest Rate Modeling. Atlantic Financial Press,
  2010.
\newblock ISBN 9780984422111.

\bibitem[Antonov and Spector(2011)]{AntonovSpector2011}
A.~Antonov and M.~Spector.
\newblock General short-rate analytics.
\newblock \emph{Risk}, pages 66--71, 2011.

\bibitem[Asvestas et~al.(2014)Asvestas, Sifalakis, Papadopoulou, and
  Saridakis]{Asvestas2014}
M~Asvestas, A.G Sifalakis, E.P Papadopoulou, and Y.G Saridakis.
\newblock Fokas method for a multi-domain linear reaction-diffusion equation
  with discontinuous diffusivity.
\newblock \emph{Journal of Physics: Conference Series}, 490\penalty0 (012143),
  2014.

\bibitem[Bacaer(2011)]{bacaer2011}
N.~Bacaer.
\newblock \emph{A short history of mathematical population dynamics},
  chapter~6, pages 35--39.
\newblock Springer-Verlag, London, 2011.
\newblock ISBN 978-0-85729-114-1.

\bibitem[Black and Karasinski(1991)]{BK1991}
F.~Black and P.~Karasinski.
\newblock Bond and option pricing when short rates are lognormal.
\newblock \emph{Financial Analysts Journal}, pages 52--59, 1991.

\bibitem[Brigo and Mercurio(2006)]{BM2006}
D.~Brigo and F.~Mercurio.
\newblock \emph{{Interest Rate Models -- Theory and Practice with Smile,
  Inflation and Credit}}.
\newblock Springer Verlag, 2nd edition, 2006.

\bibitem[Capriotti and Stehlikova(2014)]{Capriotti2014}
L.~Capriotti and B.~Stehlikova.
\newblock {An Effective Approximation for Zero-Coupon Bonds and Arrow-Debreu
  Prices in the Black-Karasinski Model}.
\newblock \emph{International Journal of Theoretical and Applied Finance},
  17\penalty0 (6):\penalty0 1650017, 2014.

\bibitem[Carr and March(2018)]{CarrMarch2018}
E.J. Carr and N.G. March.
\newblock Semi-analytical solution of multilayer diffusion problems with
  time-varying boundary conditions and general interface conditions.
\newblock \emph{Applied Mathematics and Computation}, 333\penalty0
  (15):\penalty0 286--303, 2018.

\bibitem[Carr and Itkin(2019)]{GLVG}
P.~Carr and A.~Itkin.
\newblock Geometric local variance gamma model.
\newblock 27\penalty0 (2):\penalty0 7--30, 2019.

\bibitem[Carr and Itkin(2020)]{ELVG}
P.~Carr and A.~Itkin.
\newblock An expanded local variance gamma model.
\newblock 2 2020.
\newblock \doi{10.1007/s10614-020-10000-w}.

\bibitem[Carr and Itkin(2021)]{CarrItkin2020jd}
P.~Carr and A.~Itkin.
\newblock {Semi-closed form solutions for barrier and American options written
  on a time-dependent Ornstein Uhlenbeck process}.
\newblock \emph{Journal of Derivatives}, Fall, 2021.

\bibitem[Carr and Nadtochiy(2017)]{CarrNadtochiy2017}
P.~Carr and S.~Nadtochiy.
\newblock {Local Variance Gamma} and explicit calibration to option prices.
\newblock \emph{Mathematical Finance}, 27\penalty0 (1):\penalty0 151--193,
  2017.

\bibitem[Carr et~al.(2020)Carr, Itkin, and Muravey]{CarrItkinMuravey2020}
P.~Carr, A.~Itkin, and D.~Muravey.
\newblock Semi-closed form prices of barrier options in the time-dependent cev
  and cir models.
\newblock \emph{Journal of Derivatives}, 28\penalty0 (1):\penalty0 26--50,
  2020.

\bibitem[Craddock(2009)]{Craddock2009}
M.~Craddock.
\newblock Fundamental solutions, transition densities and the integration of
  {Lie} symmetries.
\newblock \emph{Journal of Differential Equations}, 246:\penalty0 2538--2560,
  2009.

\bibitem[Dias()]{Dias2014}
C.J. Dias.
\newblock A method of recursive images to solve transient heat diffusionin
  multilayer materials.
\newblock 85:\penalty0 1075--1083.

\bibitem[Dupire(1994)]{Dupire:94}
Bruno Dupire.
\newblock Pricing with a smile.
\newblock \emph{Risk}, 7:\penalty0 18--20, 1994.

\bibitem[Fasshauer et~al.(2004)Fasshauer, Khaliq, and Voss]{Fasshauer2}
G.~E. Fasshauer, A.~Q.~M. Khaliq, and D.~A. Voss.
\newblock Using meshfree approximation for multi-asset {A}merican option
  problems.
\newblock \emph{J. Chinese Inst. Engrs.}, 27\penalty0 (4):\penalty0 563--571,
  2004.

\bibitem[Giet et~al.(2015)Giet, Vallois, and Wantz-Mezieres]{Logistic2015}
J.S. Giet, P.~Vallois, and S.~Wantz-Mezieres.
\newblock The logistic sde.
\newblock \emph{Theory of Stochastic Processes}, 20\penalty0 (36):\penalty0
  28--62, 2015.

\bibitem[Hon and Mao(1999)]{YCHon3}
Y.~C. Hon and X.~Z. Mao.
\newblock A radial basis function method for solving options pricing model.
\newblock \emph{Financial Engineering}, 8\penalty0 (1):\penalty0 31--49, 1999.

\bibitem[Horvath et~al.(2017)Horvath, Jacquier, and Turfus]{Horvath2017}
B~Horvath, A.~Jacquier, and C.~Turfus.
\newblock Analytic option prices for the black-karasinski short rate model,
  2017.
\newblock URL
  \url{https://papers.ssrn.com/sol3/papers.cfm?abstract_id=3253833}.
\newblock SSRN: 3253833.

\bibitem[Hull(2011)]{Hull2011}
J.C. Hull.
\newblock \emph{Options, Futures, and Other Derivatives}.
\newblock Prentice Hall, 8rd edition, 2011.

\bibitem[Itkin(2017)]{ItkinBook}
A.~Itkin.
\newblock \emph{Pricing derivatives under {L{\'e}vy} models}.
\newblock Number~12 in Pseudo-Differential Operators. Birkhauser, Basel, 1
  edition, 2017.

\bibitem[Itkin(2020)]{ItkinLocalVol}
A.~Itkin.
\newblock \emph{{Fitting Local Volatility: Analytic and Numerical Approaches in
  Black-Scholes and Local Variance Gamma Models}}.
\newblock Number 11623. World Scientific Publishing Co. Pte. Ltd., 2020.

\bibitem[Itkin and Lipton(2018)]{ItkinLipton2017}
A.~Itkin and A.~Lipton.
\newblock Filling the gaps smoothly.
\newblock \emph{Journal of Computational Sciences}, 24:\penalty0 195--208,
  2018.

\bibitem[Itkin and Muravey(2020{\natexlab{a}})]{ItkinMuravey2020r}
A.~Itkin and D.~Muravey.
\newblock {Semi-closed form prices of barrier options in the Hull-White model}.
\newblock \emph{Risk}, December 2020{\natexlab{a}}.

\bibitem[Itkin and Muravey(2020{\natexlab{b}})]{ItkinMuraveyDB}
A.~Itkin and D.~Muravey.
\newblock Semi-analytic pricing of double barrier options with time-dependent
  barriers and rebates at hit, September 2020{\natexlab{b}}.
\newblock URL \url{https://arxiv.org/abs/2009.09342}.

\bibitem[Itkin et~al.(2020)Itkin, Lipton, and Muravey]{ItkinLiptonMuravey}
A.~Itkin, A.~Lipton, and D.~Muravey.
\newblock From the black-karasinski to the verhulst model to accommodate the
  unconventional fed's policy, June 2020.
\newblock URL \url{https://arxiv.org/abs/2006.11976}.

\bibitem[Kartashov(1999)]{kartashov1999}
E.~M. Kartashov.
\newblock Analytical methods for solution of non-stationary heat conductance
  boundary problems in domains with moving boundaries.
\newblock \emph{Izvestiya RAS, Energetika}, \penalty0 (5):\penalty0 133--185,
  1999.

\bibitem[Kartashov(2001)]{kartashov2001}
E.M. Kartashov.
\newblock \emph{Analytical Methods in the Theory of Heat Conduction in Solids}.
\newblock Vysshaya Shkola, Moscow, 2001.

\bibitem[Kuznetsov(2013)]{Kuznetsov2013}
A.~Kuznetsov.
\newblock On the convergence of the {Gaver-Stehfest} algorithm.
\newblock \emph{SIAM J. Numerical Analysis}, 51\penalty0 (6):\penalty0
  2984--2998, 2013.

\bibitem[Lan{\c{c}}on et~al.(2001)Lan{\c{c}}on, Batrouni, Lobry, and
  Ostrowsky]{Lan_on_2001}
P~Lan{\c{c}}on, G~Batrouni, L~Lobry, and N~Ostrowsky.
\newblock Drift without flux: Brownian walker with a space-dependent diffusion
  coefficient.
\newblock \emph{Europhysics Letters ({EPL})}, 54\penalty0 (1):\penalty0 28--34,
  2001.

\bibitem[Lejay(2006)]{Lejay2006}
A.~Lejay.
\newblock On the constructions of the skew brownian motion.
\newblock \emph{Probability Surveys}, 3:\penalty0 413--466, 2006.

\bibitem[{Lienhard IV} and {Lienhard V}(2019)]{ahtt5e}
J.H. {Lienhard IV} and J.H. {Lienhard V}.
\newblock \emph{A Heat Transfer Textbook}.
\newblock Phlogiston Press, Cambridge, MA, 5th edition, 8 2019.

\bibitem[Lipton(2001)]{Lipton2001}
A.~Lipton.
\newblock \emph{Mathematical Methods For Foreign Exchange: A Financial
  Engineer's Approach}.
\newblock World Scientific, 2001.

\bibitem[Lipton and {de Prado}(2020)]{LiptonPrado2020}
A.~Lipton and M.L. {de Prado}.
\newblock A closed-form solution for optimal mean-reverting trading strategies.
\newblock \emph{Risk}, June 2020.

\bibitem[Lipton and Kaushansky(2020{\natexlab{a}})]{LiptonKau2020-2}
A.~Lipton and V.~Kaushansky.
\newblock On the first hitting time density for a reducible diffusion process.
\newblock \emph{Quantitative Finance,}, 5, 2020{\natexlab{a}}.
\newblock published online.

\bibitem[Lipton and Kaushansky(2020{\natexlab{b}})]{LiptonKaush2020}
A.~Lipton and V.~Kaushansky.
\newblock On three important problems in mathematical finance.
\newblock \emph{The Journal of Derivatives. Special Issue}, 28\penalty0 (2),
  2020{\natexlab{b}}.

\bibitem[Lipton and Sepp(2011)]{LiptonSepp2011iv}
A.~Lipton and A.~Sepp.
\newblock Filling the gaps.
\newblock \emph{Risk Magazine}, pages 86--91, 10 2011.

\bibitem[Mumford et~al.(1983)Mumford, Nori, Previato, and
  Stillman]{mumford1983tata}
D.~Mumford, C.~Musiliand~M. Nori, E.~Previato, and M.~Stillman.
\newblock \emph{Tata Lectures on Theta}.
\newblock Progress in Mathematics. Birkh{\"a}user Boston, 1983.
\newblock ISBN 9780817631093.

\bibitem[Oleinik and Radkevich(1973)]{OleinikRadkevich73}
O.~A. Oleinik and E.~V. Radkevich.
\newblock \emph{Second order equations with non-negative characteristic form}.
\newblock Kluwer Academic Publishers, 1973.

\bibitem[Pettersson et~al.(2008)Pettersson, Larsson, Marcusson, and
  Persson]{Pettersson}
Ulrika Pettersson, Elisabeth Larsson, Gunnar Marcusson, and Jonas Persson.
\newblock Improved radial basis function methods for multi-dimensional option
  pricing.
\newblock \emph{J. Comput. Appl. Math.}, 222\penalty0 (1):\penalty0 82--93,
  2008.
\newblock \doi{10.1016/j.cam.2007.10.038}.

\bibitem[Polyanin(2002)]{Polyanin2002}
A.D. Polyanin.
\newblock \emph{Handbook of linear partial differential equations for engineers
  and scientists}.
\newblock Chapman \& Hall/CRC, 2002.

\bibitem[Pontrelli et~al.(2016)Pontrelli, Lauricella, Ferreira, and
  Pena]{Pontrelli2016}
G.~Pontrelli, M.~Lauricella, J.A. Ferreira, and G.~Pena.
\newblock Iontophoretic transdermal drug delivery: A multi-layered approach.
\newblock \emph{Mathematical Medicine and Biology}, 00:\penalty0 1--18, 2016.

\bibitem[Tikhonov and Samarskii(1963)]{TS1963}
A.N. Tikhonov and A.A. Samarskii.
\newblock \emph{Equations of mathematical physics}.
\newblock Pergamon Press, Oxford, 1963.

\bibitem[Turfus(2020)]{Turfus2020}
C.~Turfus.
\newblock Analytic swaption pricing in the black-karasinski model, February
  2020.
\newblock URL
  \url{https://papers.ssrn.com/sol3/papers.cfm?abstract_id=3253866}.
\newblock SSRN: 3253866.

\bibitem[Verhulst(1838)]{Verhulst}
P.F. Verhulst.
\newblock Notice sur la loi que la population suit dans son accroisseement.
\newblock \emph{Correspondance mathematique et physique}, 10:\penalty0
  113--121, 1838.

\end{thebibliography}

\vspace{0.4in}
\appendixpage
\appendix
\numberwithin{equation}{section}
\setcounter{equation}{0}

\section{Transformation of a non-divergent heat equation to a divergent form} \label{divA}

Consider the PDE in \eqref{DupireProblem1} which is a divergent form of the heat equation
\begin{align} \label{DP1}
\fp{U(t,x)}{t} &= \fp{}{x}\left(\Xi^2(x) \fp{U(t,x)}{x}\right).
\end{align}
In this Section we show how to transform it to a non-divergent form as in \eqref{DupireProblem} when the external boundaries are constant, i.e. $y_0(t) = \chi^-(t) = const, \ y_N(t) = \chi^+(t) = const$.  We start with making a change of variables $x \mapsto z = f(x)$ with
\begin{equation}
f(x) = c_1 + c_2 \int_0^x \frac{1}{\Xi^2(k)} dk,
\end{equation}
\noindent where $c_1, c_2$ are some constants. This transformation reduces \eqref{DupireProblem1} to
\begin{align} \label{DP2}
\fp{U(t,z)}{t} &= \sigma^2(z) \sop{U(t,z)}{z}, \\
\sigma(z) &= \frac{\Xi(x(z))}{x'(z)} = \frac{c_1}{\Xi(x(z))}. \nonumber
\end{align}

The \eqref{DP2} is a non-divergent form of the heat equation. The only thing which remains to be done is finding the dependence $x(z)$. Obviously, it solves the equation
\begin{equation}
z = f(x) = c_2 + c_1 \int_0^x \frac{1}{\Xi^2(k)} dk.
\end{equation}
Given $\Xi(x)$, it can be solved either numerically (so this dependence can be precomputed), or in some cases analytically. As an example, assume that $\Xi(x) = e^{-a x}, \ a = const \ne 0$, and also let $c_2 = 0$. Then,
\begin{align}
x &= \frac{1}{a} \log\left(1 + \frac{a}{c_1}z \right), \\
\sigma^2(z) &= c_1(c_1 + a z). \nonumber
\end{align}

Reverting these steps, we obtain the inverse transformation from a non-divergent heat equation to a divergent one.

Also, the second continuity condition for \eqref{DP1} (an equality of fluxes over the boundary) is given by \eqref{matching_layers} which, by using our notation in this Section, can be re-written as
\begin{align} \label{matching_layers1}
\Xi_i^2(\yii) \fp{U_i}{x}\Bigg|_{x = \yii} &= \Xi_{i+1}^2(\yii) \fp{ U_{i+1}}{x}\Bigg|_{x = \yii}, \quad i = 1,\ldots,N-1.
\end{align}

Using \eqref{DP2} this can be transformed to
\begin{align} \label{matching_layers2}
\fp{U_i}{z}  \Bigg|_{z = z(y_{i+1})} &= \fp{ U_{i+1}}{z}\Bigg|_{z = z(y_{i+1})}, \quad i = 1,\ldots,N-1, \\
z(y_{i+1}) &= c_2 + c_1 \int_0^{y_{i+1}} \frac{1}{\Xi^2(k)} dk = c_2 + \frac{1}{c_1}
\int_0^{y_{i+1}} \sigma^2(k) dk. \nonumber
\end{align}
This is the continuity condition for \eqref{DP2}.

\section{Multilayer method for time-inhomogeneous coefficients and the domain} \label{App2}

In this section we generalize the ML method to the case $\sigma = \sigma(\tau,x)$. We again consider the initial-boundary  problem \eqref{L_Cauchy_problem} for the differential operator $\LL_i$ of the form \eqref{L_def} where now each operator $\LL_i$ reads
\begin{equation}  \label{LLi_def_TD}
\LL_i =  - \fp{}{\tau} + \sigma^2_i(\tau,x)\sop{}{x}.
\end{equation}
As before, we look for the solution of the problem \eqref{L_Cauchy_problem} in the form \eqref{u_sum_repr} such that the conditions \eqref{matching_layers} still hold. Our goal is to show that under certain assumptions the problem \eqref{L_Cauchy_problem} with time and space dependent volatility $\sigma(\tau,x)$ can be reduced to the corresponding time-homogeneous problem.

Suppose that for each sub-domain $\Omega_i$ we can construct a map $\MI$ transforming \eqref{LLi_def_TD} into PDE of the form \eqref{LLi_def}. Then the solution of the transformed PDE can be represented in the form of the heat potential \eqref{u_i_HP_repres}, and then transformed back by inverting the map $\MI$. More precisely, consider a collection of maps $\left\{\MI\right\}_{i =1}^{N}$ acting on triplets $\left(\tau, x, u_i(\tau,x) \right)$
\begin{equation}
\label{MI_def}
\left(\tau,x, u_i(\tau,x) \right) \xmapsto{\MI} \left( \TI(\tau), \XI(\tau,x), \UI(\TI, \XI) \right), \quad \TI(0) =0,
\end{equation}
\noindent such that the function $\UI(\TI, \XI)$ solves the following PDE with time-independent coefficients
\begin{equation}
\label{U_PDE}
- \fp{\UI}{\TI} + A^2(\XI) \sop{\UI}{\XI} = 0.
\end{equation}

Also, let us denote the inverse map as $\Upsilon_i(\tau,x)$, such that the following representation holds
\begin{equation}
\label{U2u}
u_i(\tau,x) = \Upsilon_i(\tau,x) \UI(\XI(\tau,x), \TI(\tau)).
\end{equation}

The map $\MI$ transforms the sub-domain $\Omega_i$ to the sub-domain $\Xi_i$
\begin{equation*}
\Xi_i \,: \, \left[\Yi^{-}(\TI), \,\Yi^{+}(\TI) \right] \times \mathbb{R}_{+}
\end{equation*}
\noindent bounded by the curves $\Yi^{-}(\TI)$ and $\Yi^{+}(\TI)$ which are defined as
\begin{equation*}
\Yi^{-}(\TI) = \XI(\lambda_i(\TI),\yi(\lambda_i(\TI))),
\quad \Yi^{+}(\TI) = \XI(\lambda_i(\TI),\yii(\lambda_i(\TI))).
\end{equation*}
Here $\lambda_i$ is the inverse map $\TI^{-1}$, i.e. $\lambda_i(\TI) =\tau(\TI) =\TI^{-1}$. Since the new time variables $\TI$ are different for each layer $\Xi_i$ the transformed boundaries are different as well, i.e., $\Yi^{+}(\TI) \neq \Y_{i+1}^{-}(\T_{i+1})$. Also, the initial value function $f(x)$ is transformed to the function $\FI$
\begin{equation*}
f(x) \xrightarrow{\MI} \FI(\XI), \qquad  \FI(\XI) = f(\eta_i\left(\XI\right)) / \Upsilon_i(\eta_i\left(\XI\right)),
\end{equation*}
\noindent where $\eta_i(x)$ solves the equation
\begin{equation*}
\eta_i(x) :  \XI(0,\eta_i(x)) = x.
\end{equation*}

Since the equations \eqref{U_PDE} are time-homogeneous, we can represent their solutions in the form of \eqref{u_i_HP_repres}
\begin{align} \label{U_transfromi_HP_repres}
\UI(\TI,\XI) &= \int_0^{\TI} \Bigg\{ \Phi_i(k) \fp{G(\XI, \xi, \TI - k)}{\xi}\Bigg|_{\xi = \Yi^{-}(k)} + \Psi_i(k) \fp{G(\XI,\xi, \TI - k)}{\xi}\Bigg|_{\xi = \Yi^+(k)} \Bigg\}dk.
\end{align}
Then making the inversion in \eqref{U2u}, applying the chain rule
\begin{equation*}
\fp{u_i}{x} = \UI(\TI(\tau),\XI(\tau,x)) \fp{\Upsilon(\tau,x)}{x} + \Upsilon(\tau,x)  \fp{\XI(\tau,x)}{x} \fp{\UI(\TI(\tau),\X)}{\X}\bigg|_{\X = \XI(\tau,x)},
\end{equation*}
\noindent and taking into account the discontinuity of the layer potentials on the boundaries, we arrive at the system of Volterra equations in \eqref{matching_layers}.

The map \eqref{MI_def} can be explicitly found via two different approaches. The first is by application of Lie symmetry analysis. It is well known, that if \eqref{U_PDE} has six or four independent groups of symmetries, it can be reduced to the heat or Bessel PDE, see \citep{Craddock2009}.

Another method is based on the theory of diffusion processes. Since any PDE of the form \eqref{LLi_def_TD} and \eqref{U_PDE} can be associated with some diffusion process, say $X = \left\{X_t, t \geq 0 \right\}$ for \eqref{LLi_def_TD}  and $Y = \left\{Y_t, t \geq 0 \right\}$ for \eqref{U_PDE}, the map in \eqref{MI_def} can be found via reduction methods, see \citep{LiptonKau2020-2} and references therein. The terms $\T(\tau),\XI(\tau,x) $ and $\Upsilon_i(\tau,x)$ are interpreted as a scale, time and measure changes.

\section{Coefficients of \eqref{Msystem}} \label{App3}

By using the definitions of coefficients of \eqref{Msystem} given in \eqref{etaDef} and \eqref{eta_def_Theta} and tables of Laplace transforms we find

\begin{align}
\La(\eta_{i}^{even}) &= \La \left\{\sum_{n =-\infty}^{\infty} \frac{e^{-\frac{(2 n l_i)^2}{4\sigma_i^2 t}}}{\sigma_i \sqrt{\pi t}}\right\}
= \frac{1}{\sigma_i \sqrt{\lambda}} \sum_{n =-\infty}^{\infty} e^{-\frac{\sqrt{\lambda} |2 n l_i|}{\sigma_i}}
= \frac{1}{\sigma_i \sqrt{\lambda}} \left( 1 + 2 \sum_{n = 1}^\infty e^{-\frac{2\sqrt{\lambda} n l_i}{\sigma_i}} \right) \\
&= \frac{1}{\sigma_i \sqrt{\lambda}} \left( 1 + \frac{2 e^{-\frac{2\sqrt{\lambda} l_i}{\sigma_i}}}{1 - e^{-\frac{2\sqrt{\lambda} l_i}{\sigma_i}}}\right)
= \frac{1}{\sigma_i \sqrt{\lambda}} \coth\left( \frac{\sqrt{\lambda} l_i}{\sigma_i}\right), \nonumber \\
\La(\eta_{i}^{odd}) &= \La \left\{\sum_{n =-\infty}^{\infty} \frac{e^{-\frac{((2 n+1) l_i)^2}{4\sigma_i^2 t}}}{\sigma_i \sqrt{\pi t}}\right\}
= \frac{1}{\sigma_i \sqrt{\lambda}} \sum_{n =-\infty}^{\infty} e^{-\frac{\sqrt{\lambda} |(2 n +1)l_i|}{\sigma_i}}
= \frac{2 e^{-\frac{\sqrt{\lambda} n l_i}{\sigma_i}}}{\sigma_i \sqrt{\lambda}} \sum_{n = 0}^\infty e^{-\frac{2\sqrt{\lambda} n l_i}{\sigma_i}} \nonumber \\
&= \frac{1}{\sigma_i \sqrt{\lambda}} \frac{2 e^{-\frac{\sqrt{\lambda} l_i}{\sigma_i}}}{1 - e^{-\frac{2 \sqrt{\lambda} l_i}{\sigma_i}}}
= \frac{1}{\sigma_i \sqrt{\lambda}} \frac{1}{\sinh\left( \frac{\sqrt{\lambda} l_i}{\sigma_i}\right)}, \nonumber \\
\La(\upsilon^-(t |x_0, 0)) &= -\frac{1}{\sigma_j^2} \sum_{n = -\infty}^{\infty} \left( y_j -x_0 +2 n l_i \right) e^{-\frac{\sqrt{\lambda}}{\sigma_j} |y_j -x_0 + 2 n l_j|} \nonumber \\
&=  -\frac{1}{\sigma_j^2} \sum_{n = 1}^{\infty} e^{-\frac{\sqrt{\lambda}}{\sigma_j} (y_j -x_0) - \frac{\sqrt{\lambda}}{\sigma_j} 2 n l_j} + \frac{1}{\sigma_j^2} \sum_{n=0}^{\infty} e^{\frac{\sqrt{\lambda}}{\sigma_j} (y_j -x_0) - \frac{\sqrt{\lambda}}{\sigma_j} 2 n l_j} \nonumber \\
&= \frac{1}{\sigma_j^2} \left[ e^{\frac{\sqrt{\lambda}}{\sigma_j}(y_j - x_0)} -
e^{-\frac{\sqrt{\lambda}}{\sigma_j}(y_j - x_0) - 2 \frac{\sqrt{\lambda}}{\sigma_j} l_j} \right]
\sum_{n = 0}^{\infty}e^{- \frac{\sqrt{\lambda}}{\sigma_j} 2 n l_j} \nonumber \\
&= \frac{1}{\sigma_j^2} \frac{ e^{\frac{\sqrt{\lambda}}{\sigma_j}(y_j - x_0)} -
e^{-\frac{\sqrt{\lambda}}{\sigma_j}(y_j - x_0) - 2 \frac{\sqrt{\lambda}}{\sigma_j} l_j} } {1 - e^{- \frac{2\sqrt{\lambda}}{\sigma_j} l_j}}
= \frac{1}{\sigma_j^2} \frac{\sinh\left(\frac{(y_j - x_0 + l)\sqrt{\lambda}}{\sigma_j}\right)} {\sinh\left( \frac{l_j \sqrt{\lambda}}{\sigma_j} \right)}
=  \frac{1}{\sigma_j^2} \frac{\sinh\left(\frac{(y_{j+1} - x_0)\sqrt{\lambda}}{\sigma_j}\right)} {\sinh\left( \frac{l_j \sqrt{\lambda}}{\sigma_j} \right)} , \nonumber \\
\La(\upsilon^+(t |x_0, 0)) &= -\frac{1}{\sigma_j^2} \sum_{n = -\infty}^{\infty} \left( y_{j+1} -x_0 +2 n l_i \right) e^{-\frac{\sqrt{\lambda}}{\sigma_j} |y_{j+1} -x_0 + 2 n l_j|} \nonumber \\
&=  -\frac{1}{\sigma_j^2} \sum_{n = 0}^{\infty} e^{-\frac{\sqrt{\lambda}}{\sigma_j} (y_{j+1} -x_0) - \frac{\sqrt{\lambda}}{\sigma_j} 2 n l_j} + \frac{1}{\sigma_j^2} \sum_{n = 1}^{\infty}e^{\frac{\sqrt{\lambda}}{\sigma_j} (y_{j+1} -x_0) - \frac{\sqrt{\lambda}}{\sigma_j} 2 n l_j} \nonumber \\
&= \frac{1}{\sigma_j^2} \left[ e^{\frac{\sqrt{\lambda}}{\sigma_j}(y_{j+1} - x_0) - 2  \frac{\sqrt{\lambda}}{\sigma_j} l_j } - e^{-\frac{\sqrt{\lambda}}{\sigma_j}(y_{j+1} - x_0) } \right]
\sum_{n = 0}^{\infty}e^{- \frac{\sqrt{\lambda}}{\sigma_j} 2 n l_j} \nonumber \\
&= \frac{1}{\sigma_j^2} \frac{ e^{\frac{\sqrt{\lambda}}{\sigma_j}(y_{j+1} - x_0)} -
	e^{-\frac{\sqrt{\lambda}}{\sigma_j}(y_{j+1} - x_0) - 2 \frac{\sqrt{\lambda}}{\sigma_j} l_j} } {1 - e^{- \frac{2\sqrt{\lambda}}{\sigma_j} l_j}}
= \frac{1}{\sigma_j^2} \frac{\sinh\left(\frac{(y_{j+1} - x_0 - l)\sqrt{\lambda}}{\sigma_j}\right)} {\sinh\left( \frac{l_j \sqrt{\lambda}}{\sigma_j} \right)}
=  - \frac{1}{\sigma_j^2} \frac{\sinh\left(\frac{x_0 - y_{j})\sqrt{\lambda}}{\sigma_j}\right)} {\sinh\left( \frac{l_j \sqrt{\lambda}}{\sigma_j} \right)}. \nonumber
\end{align}

\end{document}